\documentclass{article}

\usepackage[version=4]{mhchem}
\usepackage{siunitx}
\usepackage{mathptmx}
\usepackage[scaled=0.92]{helvet}
\usepackage{microtype}
\usepackage[british]{babel}
\usepackage{graphicx}
\usepackage[a4paper, margin=2cm, bottom=3cm]{geometry}
\usepackage[format=hang, labelfont=bf]{caption}
\usepackage[colorlinks, citecolor=black]{hyperref}

\usepackage{rsc}

\DeclareSIUnit\angstrom{\text{\AA}}

\graphicspath{{figures/}}

\newcommand{\amsulfd}{\ce{(ND4)2SO4}}

\title{\sffamily\bfseries Origin of the Large Entropy Change in the Molecular Caloric and Ferroelectric Ammonium Sulfate}
\author{Shurong Yuan\footnotemark[2] \and %
  Bernet E. Meijer\footnotemark[2] \and %
  Guanqun Cai\footnotemark[2] \and %
  Richard J. C. Dixey \and %
  Franz Demmel \and %
  Martin T. Dove \and %
  Jiaxun Liu \and %
  Helen Y. Playford \and %
  Helen C. Walker\footnotemark[1] \and %
  Anthony E. Phillips\footnotemark[1]} 
\date{}

\begin{document}
\maketitle

\renewcommand{\thefootnote}{\fnsymbol{footnote}}
\footnotetext[1]{a.e.phillips@qmul.ac.uk, helen.c.walker@stfc.ac.uk}
\footnotetext[2]{These three authors contributed equally as joint first co-authors.}

{\footnotesize\sffamily\setlength{\parskip}{0.4\baselineskip}
  \noindent S. Yuan, B. E. Meijer, G. Cai, R. J. C. Dixey, M. T. Dove, J. Liu, A. E. Phillips\\
  School of Physical and Chemical Sciences, Queen Mary University of London, Mile End Rd, London E1 4NS, U.K.

  \noindent F. Demmel, H. Y. Playford, H. C. Walker\\
  ISIS Neutron and Muon Source, Rutherford Appleton Laboratory, Didcot OX11 0QX, U.K.

\noindent M. T. Dove\\School of Computer Sciences, Sichuan University, No 24 South Section 1, Yihuan Road, Chengdu, 610065, China\\
Department of Physics, Wuhan University of Technology, 205 Luoshi Road, Hongshan district, Wuhan, Hubei, 430070, China\\
School of Mechanical Engineering, Dongguan University of Technology, 1st Daxue Road, Songshan Lake, Dongguan, Guangdong 523000, China

}

\vspace{16pt}


\vspace{16pt}

\begin{abstract}\normalsize\noindent
  The deceptively simple inorganic salt ammonium sulfate undergoes a ferroelectric phase transition associated with a very large entropy change and both electrocaloric and barocaloric functionality. While the structural origins of the electrical polarisation are now well established, those of the entropy change have been controversial for over fifty years. This question is resolved here using a combination of DFT phonon calculations with inelastic neutron scattering under variable temperature and pressure, supported by complementary total and quasielastic neutron scattering experiments. A simple model of the entropy in which each molecular ion is disordered across the mirror plane in the high symmetry phase, although widely used in the literature, proves to be untenable. Instead, the entropy arises from low-frequency librations of ammonium ions in this phase, with harmonic terms that are very small or even negative. These results suggest that, in the search for molecular materials with functionality derived from large entropy changes, vibrational entropy arising from broad energy minima is likely to be just as important as configurational entropy arising from crystallographic disorder.
\end{abstract}

\vspace{24pt}

\section{Introduction}
Phase transitions in orientationally disordered crystals have been known for many years\cite{guthrie_observations_1961, clark_pre-melting_1974}, but have recently seen a dramatic resurgence of interest because of their promising caloric and electrical properties\cite{das_harnessing_2020}. If the entropy associated with the order-disorder transition couples to an external field, such as electric field (the electrocaloric effect) or pressure (the barocaloric effect), then this coupling can be used to produce a solid-state heat pump; these technologies hold great promise to replace environmentally damaging vapour-compression refrigerants\cite{moya2014caloric}. Likewise, if a disordered component has an electrical dipole moment, then dynamically disordered phases are likely to have large dielectric constants and good ionic conductivity, while ordered phases may show macroscopic polarisation and hence pyro- or ferroelectricity.\cite{macfarlane_plastic_2001, hang_metalorganic_2011, zhu_organic_2019, harada_plasticferroelectric_2021}

Here we revisit one of the oldest known ferroelectrics, the inorganic salt ammonium sulfate, \ce{(NH4)2SO4}. This undergoes a first-order phase transition at \SI{223}{K} and ambient pressure, involving the loss of a mirror plane perpendicular to the crystallographic $c$ axis and, as a result, ferroelectric polarisation along this axis (Fig.~1) \cite{matthias1956ferroelectricity}.  This phase transition has been extensively studied for over 50 years \cite{unruh1970spontaneous}. 
Renewed interest has recently developed in the strong electrocaloric \cite{scott2011electrocaloric} and (inverse) barocaloric effects \cite{ASgiant}, due to the unusually high entropy of transition. The caloric behaviour is especially impressive when measured per unit mass, giving an entropy change of $\Delta S = \SI{60}{J.K^{-1}.kg^{-1}}$, since ammonium sulfate contains no heavy metals and has only a modest density of \SI{1.77}{g.cm^{-3}}. The corresponding volume change of the unit cell is \SI{-4.4}{\angstrom^3}, corresponding to $\Delta V = \SI{-4.7}{cm^3.kg^{-1}}$ \cite{ASgiant}. Since this is rather large, the gradient of the phase boundary $|\mathrm d T/\mathrm d p| = |\Delta V/\Delta S| = \SI{80}{K.GPa^{-1}}$ is also impressively steep, giving a large temperature change under modest pressure.

\begin{figure}
    \centering
    \includegraphics[width=0.5\linewidth]{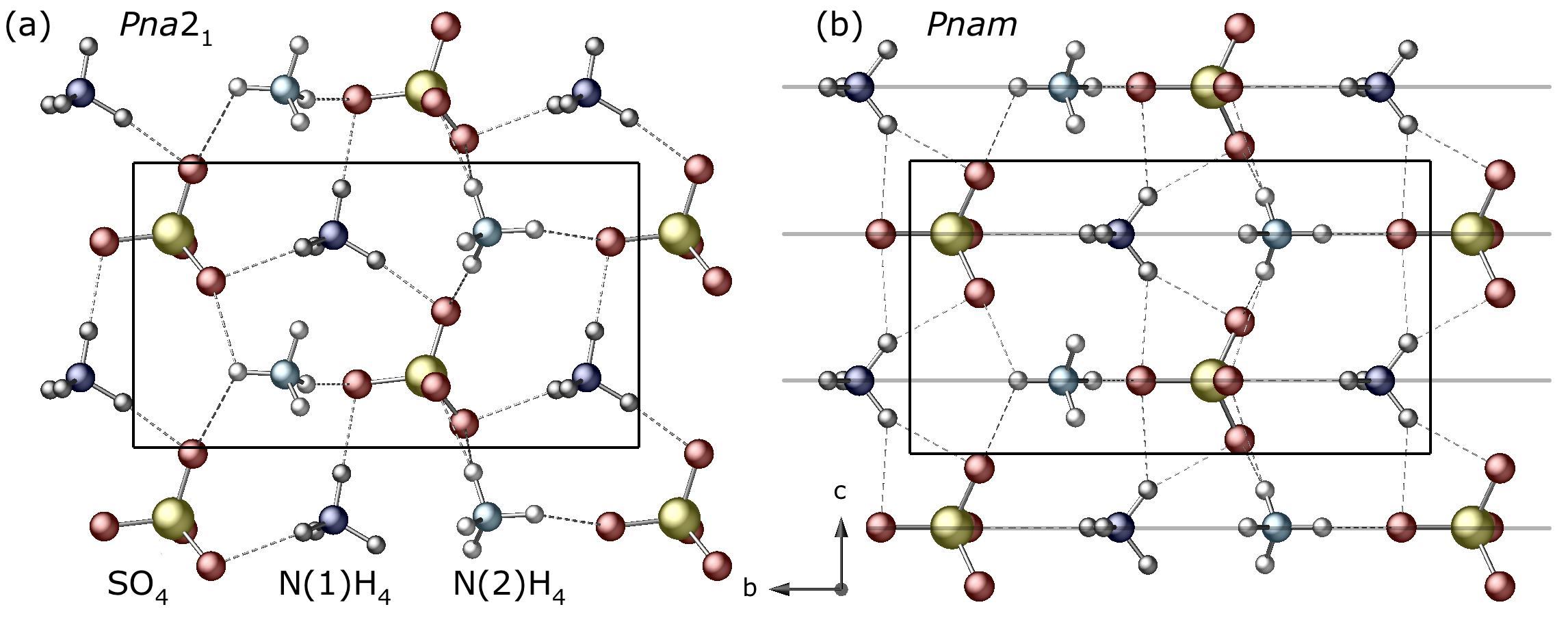}
    \caption{Structure of ammonium sulfate in the (a) low-symmetry $Pna2_1$ and (b) high-symmetry $Pnam$ phases. For clarity, only half of the molecules in the unit cell are shown, and hence not every hydrogen bond is represented in this figure. In the low-symmetry phase, the ammonium ions tilt away from the incipient mirror plane, forming fewer, shorter hydrogen bonds, while in the high-symmetry phase, the ammonium ions sit on this plane (gray lines), forming a more symmetrical arrangement of longer, weaker hydrogen bonds.}
    \label{fig:structure}
\end{figure}

There has been substantial discussion of the mechanism of the phase transition, focused mainly on the origin of the electrical polarisation and hence the ferroelectric properties.
This question appears to have been resolved by recent single-crystal diffraction measurements, demonstrating that the polarisation is due primarily to distortions of the ammonium ions. The phase transition involves cooperative rotation of the sulfate ions and a concomitant rearrangement of the \ce{NH4\bond{...}SO4} hydrogen-bonding network \cite{unmasking}. 
On the other hand, the origin of the caloric properties -- the \emph{entropy} change at the phase transition -- remains uncertain. The entropy change was initially explained in terms of an order-disorder model in which the three tetrahedral ions (two ammonium ions and one sulfate ion) in each formula unit are independent ``statistical units'', showing twofold disorder about the mirror plane above the phase transition but ordering below it \cite{oreilly_deuteron_1967}.
In this case, the Boltzmann formula for configurational entropy predicts a transition entropy of $(3\ln 2) R = \SI{17.3}{J.K^{-1}.mol^{-1}}$, in good agreement with the experimental value of \SI{17.6}{J.K^{-1}.mol^{-1}} \cite{hoshino_dielectric_1958}.
Spectroscopic data were also interpreted as favouring an order-disorder model \cite{oreilly_orderdisorder_1969, de_sousa_meneses_phase-transition_1995}.

This model, however, is unsatisfactory for two reasons. First, it
is not consistent with the substantial body of accumulated crystallographic data on this material, since 
there is almost no crystallographic evidence for a disordered structure in the paraelectric phase. One exception is an early X-ray study, which appeared to show some evidence of hydrogen disorder \cite{hasebe_studies_1981}, but this was neither reproduced by the recent X-ray measurements \cite{unmasking} nor shown in neutron experiments, which are far more sensitive to scattering from H atoms \cite{schlemper_neutrondiffraction_1966}. By contrast, crystallographic disorder is clearly visible in materials with similar structures, including \ce{(NH4)2BeF4} \cite{yamada_disorder_1985} and \ce{(N(CH3)4)2ZnCl4} \cite{hasebe_x-ray_1987}. Similarly, in vibrational spectroscopy data, an order-disorder transition would be expected to produce a sudden change in the width of relevant peaks at the phase transition, whereas Raman spectroscopy shows no such change \cite{torrie1972raman}. Second, the simple order-disorder model makes no attempt to account for dynamics, which will certainly contribute to the entropy change \cite{butler_organised_2016}. 
An alternative explanation is therefore needed for the remarkably large entropy change in ammonium sulfate. This problem was considered very recently by Born-Oppenheimer molecular dynamics simulation \cite{malec_displacive_2021} but has not yet been tackled experimentally.

To resolve this question, we have studied the local structure and dynamics of ammonium sulfate computationally, by density-functional theory calculations of the phonon spectrum; and experimentally, by inelastic, quasielastic, and total neutron scattering, as functions of temperature and pressure. Inelastic neutron scattering is sensitive to the vibrational dynamics of all atoms in the material; quasielastic neutron scattering reveals the reorientation dynamics of the ammonium ions; and total neutron scattering reflects the correlations between instantaneous atomic positions. This combination of techniques therefore provides access to all sources of our target material's entropy.

In Section~\ref{sec:config vs vib}, we show that the large entropy change in ammonium sulfate cannot meaningfully be described as configurational (\emph{i.e.}, order-disorder). It is better understood as dynamic, resulting from molecular librations that correspond to motion in shallow energy wells in configuration space. In Section~\ref{sec:atomic origins}, we investigate the atomic origins of this energy surface, demonstrating that it results from competition between two different arrangements of the hydrogen bonds between ammonium and sulfate ions.


\section{Is the entropy change configurational or vibrational?}\label{sec:config vs vib}

\subsection{Evidence for vibrational entropy}

A DFT model of each phase was constructed starting from the experimentally determined structures~(Fig.~\ref{fig:structure}). In each case, the two ammonium ions in the formula unit are crystallographically distinct, referred to as N1 and N2.  The network of hydrogen bonds around each of these ions is different between the two phases: the low-symmetry phase has fewer, shorter and therefore stronger hydrogen bonds, while the high-symmetry phase has a more symmetrical arrangement with more, longer, weaker hydrogen bonds.

Allowing the unit cell parameters to relax gave reasonable agreement with experiment in the $Pna2_1$ phase, but an unacceptably large divergence from the experimental structure in the $Pnam$ phase. In particular, there was a 9\% expansion along the $c$ axis compared to experimental data at \SI{233}{K}, which we attribute to neglect of thermal expansion, which will be significant in this material. For our purposes, however, it is sufficient to adopt the quasiharmonic approximation, fixing the cell parameters at their experimental values at temperatures of \SI{5}{K} (our own work; see below) and \SI{233}{K},\cite{Malec} respectively.

\begin{figure*}
\centering
\includegraphics[width=0.85\textwidth]{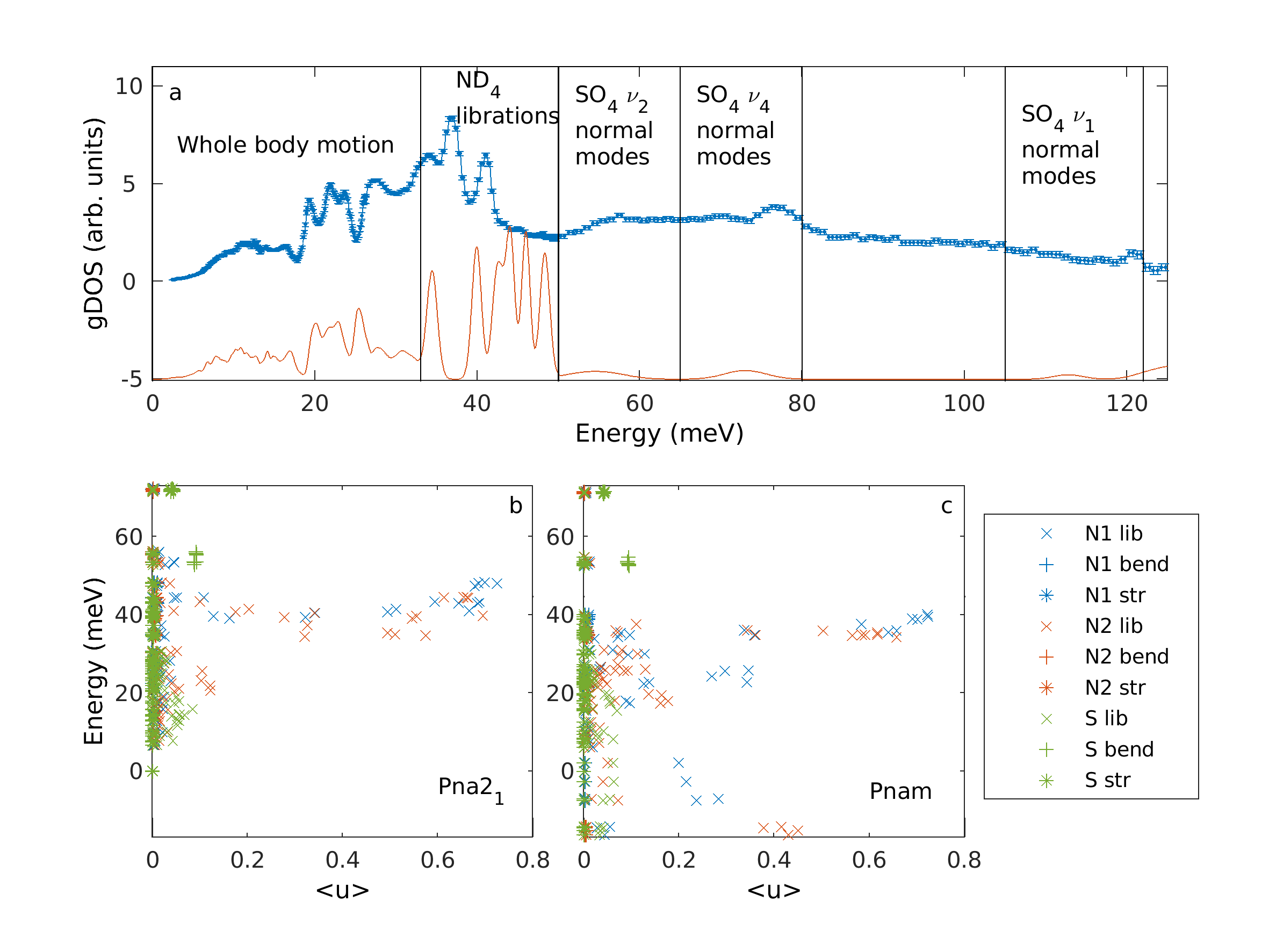}
\caption{(a) A comparison of the measured ($T=10$~K) and calculated generalised phonon density of states of \amsulfd\ in $Pna2_1$ at ambient pressure. The data were measured with four different incident energies $E_i=23, 36, 67$ and $162$~meV. The overlap between these data sets was used to estimate the background in each case and to scale the $y$-axis by an appropriate factor, allowing us to produce a
  ``continuous'' trace
  for the combined data.
  The calculated data have been broadened by a gaussian representing the instrumental resolution for the incident energies used. Specific phonon modes have been labelled on the data and simulation.
The calculated phonon modes at $\Gamma$ were separated into librations, bends and stretches of the different molecules using GASP. The energies and displacement magnitudes for the different modes in $Pna2_1$ and $Pnam$ are shown in (b) and (c).}\label{figgdosmodes}
\end{figure*}

Next, the phonon spectrum was calculated using density-functional perturbation theory. The phonon density of states is shown in Fig.~\ref{figgdosmodes}a. To characterise individual modes, the GASP algorithm\cite{Wells.2002, Wells.2004, wells_gasp_2015} was used to decompose the DFT eigenvectors at the $\Gamma$ point into translations, librations, and distortions of the two crystallographically independent ammonium ions and the sulfate ion (Fig.~\ref{figgdosmodes}b and c).\footnote{GASP, which was not primarily written with phonons in mind, refers to librations as ``rotations''; we maintain the distinction here to avoid confusion with the rotational ``hopping'' motion to which QENS measurements are sensitive.} A full comparison between these calculations and experimental data will be deferred to Section~\ref{sec:atomic origins}. For now we note that the simulation agrees well with the experimental phonon density of states from inelastic neutron scattering measurements (Fig.~\ref{figgdosmodes}a). 
There is excellent agreement up to $E=34$~meV, with the dip at $18$~meV being accurately reproduced, and only a slight shift in energy between the data and calculation for the next clear dip at $25$~meV and $24$~meV respectively. Above $34$~meV, the calculations show there are no more translational modes. The peaks in the gDOS around $40$~meV were identified as ammonium librations as opposed to translations on the basis of their negative Gr\"{u}neisen parameters (Section~\ref{sec:atomic origins}). The calculations slightly overestimate the energies corresponding to \ce{ND4} libration modes, then slightly underestimate the energies of the internal modes of both the \ce{ND4} and \ce{SO4} ions.

This analysis reveals that the key vibrational difference between the phases is the librational motion of the ammonium ions (red and blue $\times$ shapes, Fig.~\ref{figgdosmodes}b and c). In the $Pna2_1$ phase, these occur in the 34--50~meV region, where they are well separated from molecular translational modes below about 30~meV. By contrast, in $Pnam$, the librations mix with the molecular translational modes.
Even more strikingly, these decrease in frequency -- some modestly, others substantially -- such that the lowest-frequency of these librations become weakly unstable, within the harmonic approximation, in some or all of the Brillouin zone. 



\begin{figure*}
  \centering
  \includegraphics[width=\linewidth]{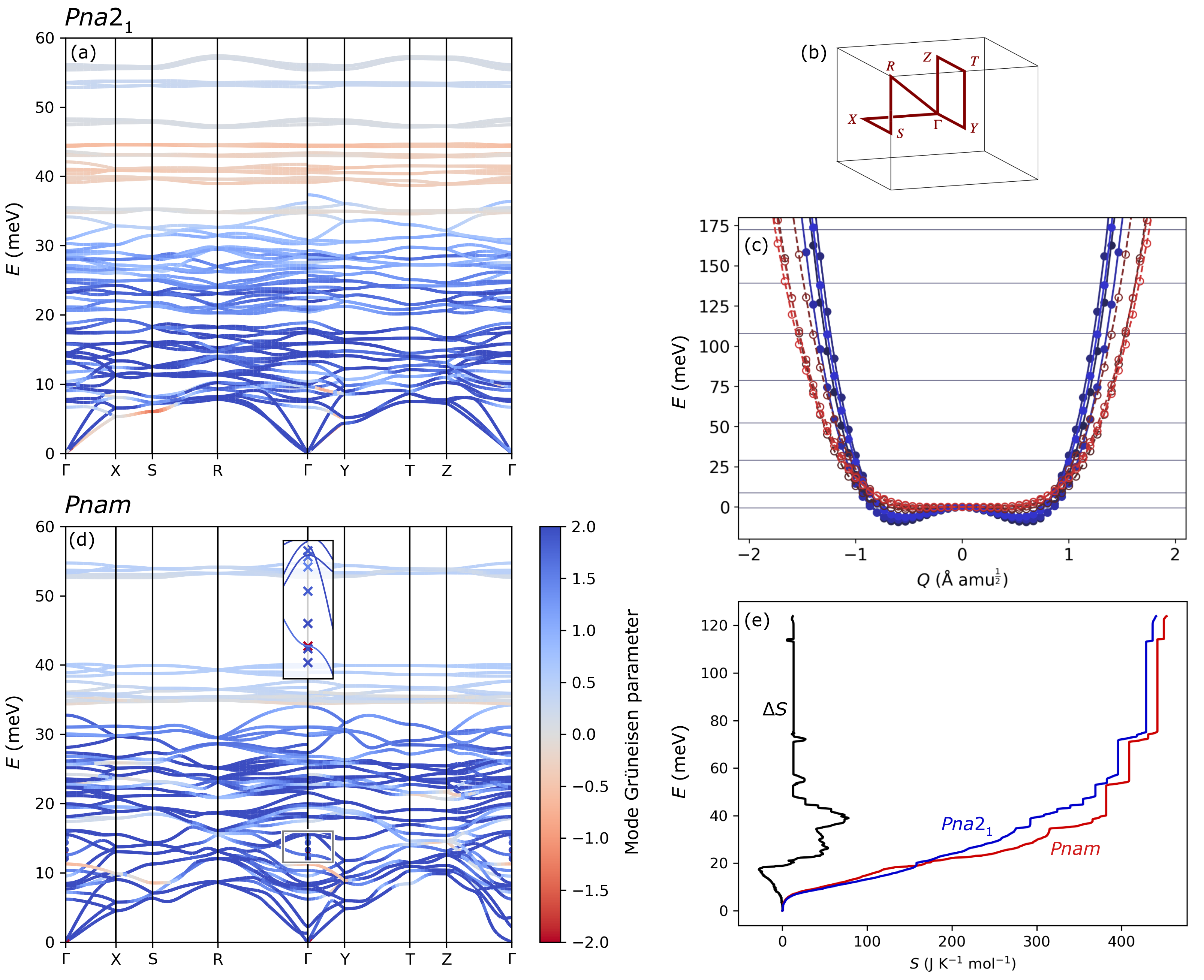}
  \caption{(a) DFT phonon dispersion curves in the $Pna2_1$ phase, following the path through the Brillouin zone shown in (b). The color of the line represents the mode Gr\"{u}neisen parameter. (c) Energies of the eight unstable modes in the $Pnam$ phase as a function of mode coordinate $Q$, calculated at $\Gamma$. There are two groups of four librations (blue solid lines and filled points: N2 librations about $b$; red dashed lines and open points: N1 librations about $a$). The energy levels associated with the lowest frequency of these modes are also shown. 
  The motions associated with these modes are shown in Fig.~S3.
  (d) DFT phonon dispersion curves in the $Pnam$ phase, following the same path and again colored to show the Gr\"uneisen parameter. The eight unstable modes have effective frequencies calculated at \SI{235}{K} at $\Gamma$ alone; these are magnified in the inset. 
  (e) Cumulative entropy in each phase, and entropy change between the phases, at 1148 K, the transition temperature determined by DFT. Modes are summed in increasing order of frequency. The entropy of transition is dominated by the increased density of states in the $Pnam$ phase from about \SIrange{20}{40}{meV}. 
  }
  \label{fig:dispersion}
\end{figure*}

The dispersion curves in the $Pna2_1$ phase are shown in Fig.~\ref{fig:dispersion}a, plotted along the path in Fig.~\ref{fig:dispersion}b. 
To account approximately for the instabilities in the $Pnam$ phase, the energy of each unstable mode was mapped as a function of the mode coordinate $Q$ at the gamma point. This revealed flat-bottomed, ``bathtub''-shaped potentials with very low harmonic terms; the energy is therefore dominated by quartic and higher-order terms (Fig.~\ref{fig:dispersion}c). Solving the Schr\"odinger equation for the resulting potentials gave a renormalised frequency that now varies with temperature \cite{skelton_anharmonicity_2016}, falling in the range \SIrange{11}{16}{meV} at $T = \SI{235}{K}$. For practical reasons, we made the strong assumption that these effective frequencies remain roughly constant across the Brillouin zone. Fig.~\ref{fig:dispersion}d shows the dispersion curve after this renormalisation; the original, showing harmonic instabilities, is given in Fig.~S1.

Finally, the entropy of transition was calculated using the standard relationship\cite{fultz_vibrational_2010}
\begin{equation}
  \label{eq:vibS}
  S_\text{vib} = k_\mathrm{B}\sum_i\left(-\ln\big(1 - \exp(-x_i)\big) + \frac{x_i}{\exp(x_i) - 1}\right), 
\end{equation}
where $x_i = \hbar\omega_i/k_\mathrm{B}T$ runs through harmonic vibrational modes with angular frequencies $\omega_i$.  This in turn allows the Helmholtz free energy $A = U - TS$ to be calculated as a function of temperature for each phase, and the phase transition temperature to be estimated by setting the free energies of each phase to be equal. (The $p\Delta V$ contribution is negligible at ambient pressure.)
This calculation gives an entropy change of \SI{13.2}{J.K^{-1}.mol^{-1}} across the phase transition, 75\% of the experimental value of \SI{17.6}{J.K^{-1}.mol^{-1}}. The phase transition temperature is predicted to be \SI{1147}{K}, substantially above the experimental value of \SI{223}{K}. The cumulative difference in entropy at the phase transition temperature is shown in Fig.~\ref{fig:dispersion}e.

We consider these values to be reasonably consistent with experiment, since the agreement is constrained by the relation $\Delta U = T\Delta S$. The DFT model gives $\Delta U = \SI{144}{meV}$ per formula unit while experimentally $T\Delta S = \SI{40}{meV}$. A 0.1~eV difference, of the order expected for DFT accuracy, requires $T$ and $\Delta S$ to differ from their experimental values by a combined factor of $3.6$. The accuracy obtained in practice is therefore as good as can be expected from this simple model. Specifically, the internal energy difference between the two phases is reproduced with only modest accuracy, which is perhaps expected given known limitations of DFT and the substantial differences in hydrogen bonding geometry between the phases. On the other hand, the difference in the shapes of the energy surfaces about the respective minima is relatively well reproduced.

To elucidate further the DFT $\Delta U$ value, we used GULP \cite{gale_gulp_2005} to calculate the Coulomb energy of a hypothetical point-charge model in which the full $+1$ charge of the ammonium ion resides on the N atom and the $-2$ of the sulfate ion on the S atom. This gives an energy difference of \SI{77.5}{meV} per formula unit purely from long-range ionic interactions, without considering changes in covalent or hydrogen bonding; this accounts for roughly half of the DFT energy difference.

One possible source of discrepancy between the experimental and calculated $\Delta S$ values is the strong approximation that the renormalised frequencies are independent of wavevector. Indeed, only a 17\% decrease in the effective frequencies of just the eight renormalised modes would be sufficient to increase the entropy change to its experimental value; this is likely to be within the error of the approximation. However, extending this calculation to consider points away from the zone center and map changes in anharmonicity across the Brillouin zone would be computationally expensive, requiring phonon calculations on large supercells.

Nonetheless, it is clear that the ammonium librational modes differ between the two phases in a way that can account for a large difference in vibrational entropy. In Section~\ref{sec:atomic origins}, we investigate further the atomic features of these modes that are responsible for this behaviour. Before proceeding to this, however, we show that, by contrast, a configurational model \emph{cannot} account for the observed entropy change.

\subsection{Evidence against configurational entropy}


Considering possible evidence for configurational entropy, we first note that the established structural model of the $Pnam$ phase is fully ordered,\cite{schlemper_neutrondiffraction_1966, unmasking} and thus has a configurational entropy of zero. Our own single-crystal X-ray diffraction data in the high-temperature phase are in good agreement with this published structure. To complement this traditional crystallographic analysis, we therefore measured the pair distribution function, which represents instantaneous local correlations between pairs of atoms and is thus sensitive to local deviations from the crystallographic average.

Total neutron scattering data were collected from a powder \amsulfd\ sample using the POLARIS diffractometer (ISIS, U.K.) at several temperatures spanning each phase. First, the crystallographic average structure in each phase was determined by  Rietveld refinement of the established, fully ordered, structural model. The refined models gave good fits to the data and agreed well with literature values, without the need to invoke crystallographic disorder. 



Next, we used reverse Monte Carlo refinement, as implemented in the RMCprofile code \cite{tucker_rmcprofile_2007}, to derive an ensemble of models, each representing an instantaneous ``snapshot'' of a $6\times 5\times 8$ supercell. In this method, the models are refined simultaneously against the pair distribution function, the scattering function, and the Bragg profile; thus they are consistent with both the local structure and the long-range average structure. Sample fits are shown in Fig.~S2; a full analysis of these data is in preparation.

\begin{figure}
  \centering
  \includegraphics[width=0.5\linewidth]{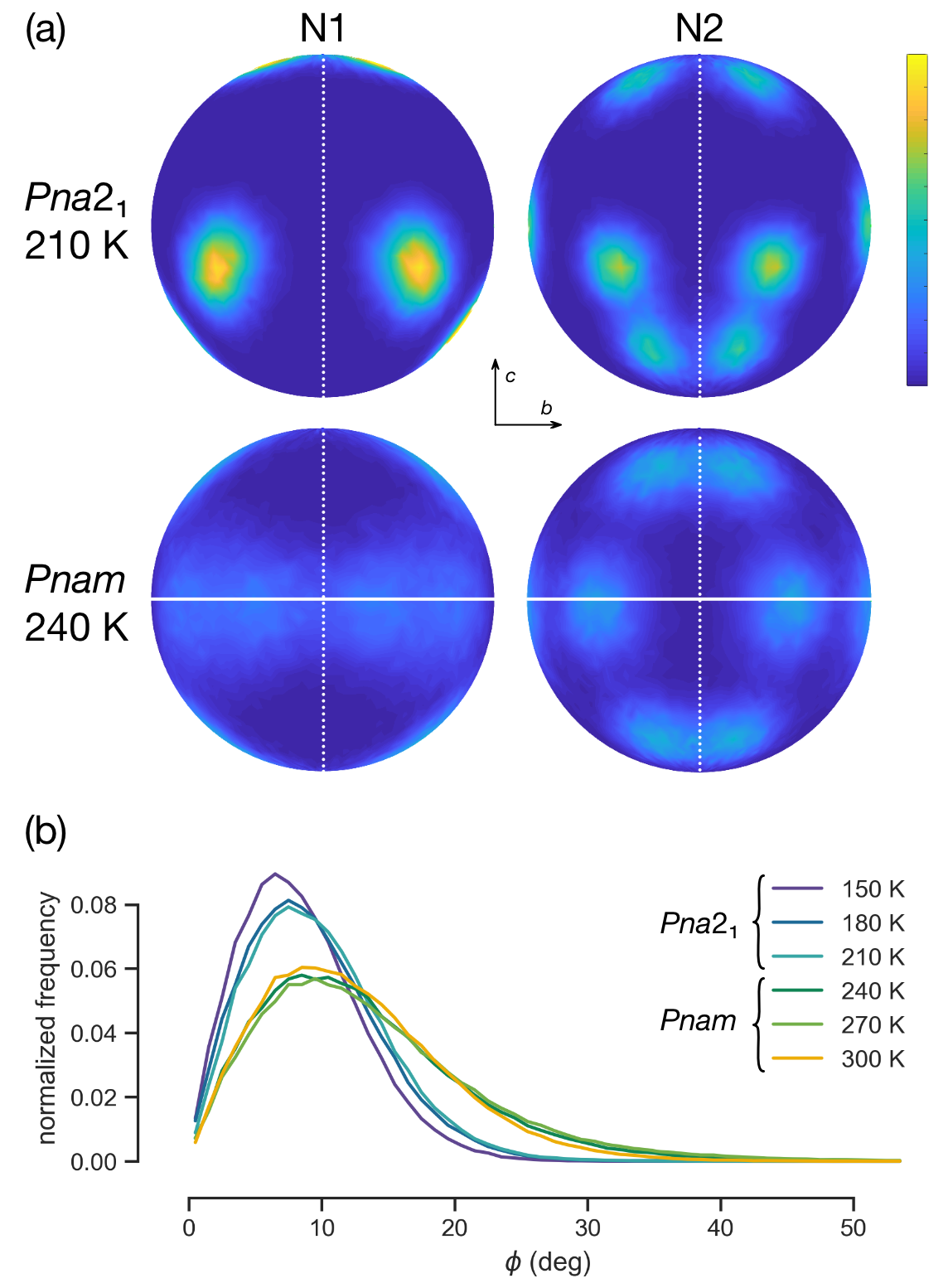}\\
  \caption{%
  (a) Spherical histograms showing the distribution of N--H bond orientations in configurations derived by RMC from total neutron scattering data, at \SI{210}{K} ($Pna2_1$) and \SI{240}{K} ($Pnam$). The diagrams are orthographic projections down the $a$ axis. The symmetries of these histograms are the respective crystal classes, $mm2$ and $mmm$: thus the vertical mirror planes (dotted line) arise from the $a$ glide plane in both phases, while the horizontal mirror plane (solid line) is the true mirror plane, present only in the high-symmetry phase. The distribution is clearly centred \emph{on} this plane, rather than having separate peaks above and below it.
  (b) Distribution of ammonium ion rotations $\phi$ from the crystallographic average position, from GASP analysis of these configurations. The distributions evolve slightly with temperature, especially in the low-symmetry phase, but differ far more dramatically between phases, again demonstrating the increased freedom of motion in the high-symmetry phase that is responsible for its increased entropy. These data have been normalised by the solid angle differential $\sin\phi\, d\phi d\theta$. In both parts, there are 9600 ammonium ion geometries, corresponding to five independent RMC runs.} 
  \label{fig:tet-rot}
\end{figure}

The directions of the N--H bonds were extracted from these models. Plotting these as a spherical histogram in orthographic projection shows isolated peaks in each phase, with no sign of the bimodal shape that would be expected from disorder about the mirror plane (Fig.~\ref{fig:tet-rot}a). However, these distributions are far broader in the high-symmetry than in the low-symmetry phase. To investigate this further, we again used the GASP algorithm to find the rotation that best related each tetrahedron in the configuration to its average crystallographic orientation. This confirmed that the ammonium ions have far more librational freedom in the high-symmetry phase (Fig.~\ref{fig:tet-rot}b). These results are thus consistent both with the established structural model and our conclusion above that ammonium librations are more thermally accessible in the high-symmetry $Pnam$ phase, but inconsistent with the simple model where the entropy of transition is purely configurational.


These results are furthermore consistent with our recent measurements of the rotational \emph{dynamics} of ammonium ions in this material using quasielastic neutron scattering.\cite{MeijerQENS}
This technique is sensitive specifically to the ammonium ions because it measures the incoherent scattering, which in this sample is dominated by \ce{^1H}; it reflects the ``hopping'' motion of these ions between energy minima corresponding to different orientations. 
We refer the interested reader to that paper for full details, but repeat here the two key results relevant to the entropy change. 

First, the proportion of \emph{elastic} scattering as a function of $Q$, known as the elastic incoherent structure factor (EISF), is determined by the \emph{geometry} of hopping; that is, by the equilibrium distribution of the moving \ce{^1H} nuclei. Hopping between two closely separated sites, on opposite sides of the mirror plane in the high-symmetry phase, would be clearly visible in this signal. However, no such hopping was observed. Again, then, these data are not consistent with the entropy change being caused by an order-disorder transition.

Second, the frequency of reorientation is indicated by the width of the characteristic Lorentzian peak in energy transfer. 
In the low-symmetry phase at \SI{200}{K}, this frequency unexpectedly increased with pressure as the phase transition was approached. 
In a simple model where the ammonium ions move in a static energy landscape, this indicates that the energy barrier to rotation is decreasing. This is consistent with the decrease in this energy barrier with \emph{temperature} reported on the basis of spin-lattice relaxation times \cite{oreilly_deuteron_1967}. Of course, in reality the surrounding ions also move, and the observed behaviour may equally be attributed to the barrier collapsing more frequently; this subtlety does not, however, affect our qualitative argument.

An increase in hopping frequency with \emph{pressure} is expected for single-atom hopping, for instance in asymmetric hydrogen bonds, since forcing two wells closer together will lower the barrier between them \cite{benoit_tunnelling_1998}. It is, however, very unusual for rotations of whole molecules such as ammonium ions, where hopping typically decreases in frequency with applied pressure \cite{kozlenko_nmr_1999}. We propose that this highly atypical behaviour reflects a flattening of the energy landscape as the phase transition is approached, both because the existing hydrogen bonds are destabilised by compression beyond their equilibrium length, and because the alternative high-symmetry arrangement with longer bonds is stabilised for the same reason. This result suggests that the ammonium \emph{librations} may be important in determining the entropy change, an idea we return to in the following section.
This change in the energy landscape is also reflected in the volume decrease across the phase transition. 
Similar behaviour was recently reported in the related material ammonium thiocyanate \cite{zhang_barocaloric_2021}.

Our results so far show that the entropy of transition is likely to be mostly vibrational rather than configurational, due to the librational motion of the ammonium ions. Next, we performed a detailed investigation to characterise why these particular vibrational modes give such a large entropy change.

\section{Atomic characterisation of the high-entropy vibrations}\label{sec:atomic origins}

\subsection{Ambient-pressure single crystal inelastic neutron scattering}

To characterise the vibrational behaviour in more detail, we collected inelastic neutron scattering from a \SI{1.2}{g} mosaic sample prepared by gluing individual \amsulfd\ crystals to aluminium plates (Fig.~S6).  Inelastic scattering techniques are an excellent probe of coherent excitations such as phonons, while neutrons show a particular benefit when it comes to the study of low-$Z$ elements, as present in ammonium sulfate. In contrast to Raman scattering, 
inelastic neutron scattering allows the dispersion of phonons to be mapped over a wide range of momentum and energy transfer space. 
In the particular case of ammonium sulfate, there is no change in systematic absences between the space groups $Pnam$ and $Pna2_1$, so we expect the changes in scattering to be more subtle than a mode simply lifting away from the elastic line. For this reason, collecting
data over a large region of 4-dimensional $\mathbf{Q}$-$E$ space is particularly important.
Indeed, 
little change in the relative intensities of the different Bragg peaks is observed in the elastic data, as expected given the subtle difference between the two known structures (Fig.~S7). 


\begin{figure*}
\centering
\includegraphics[width=\textwidth]{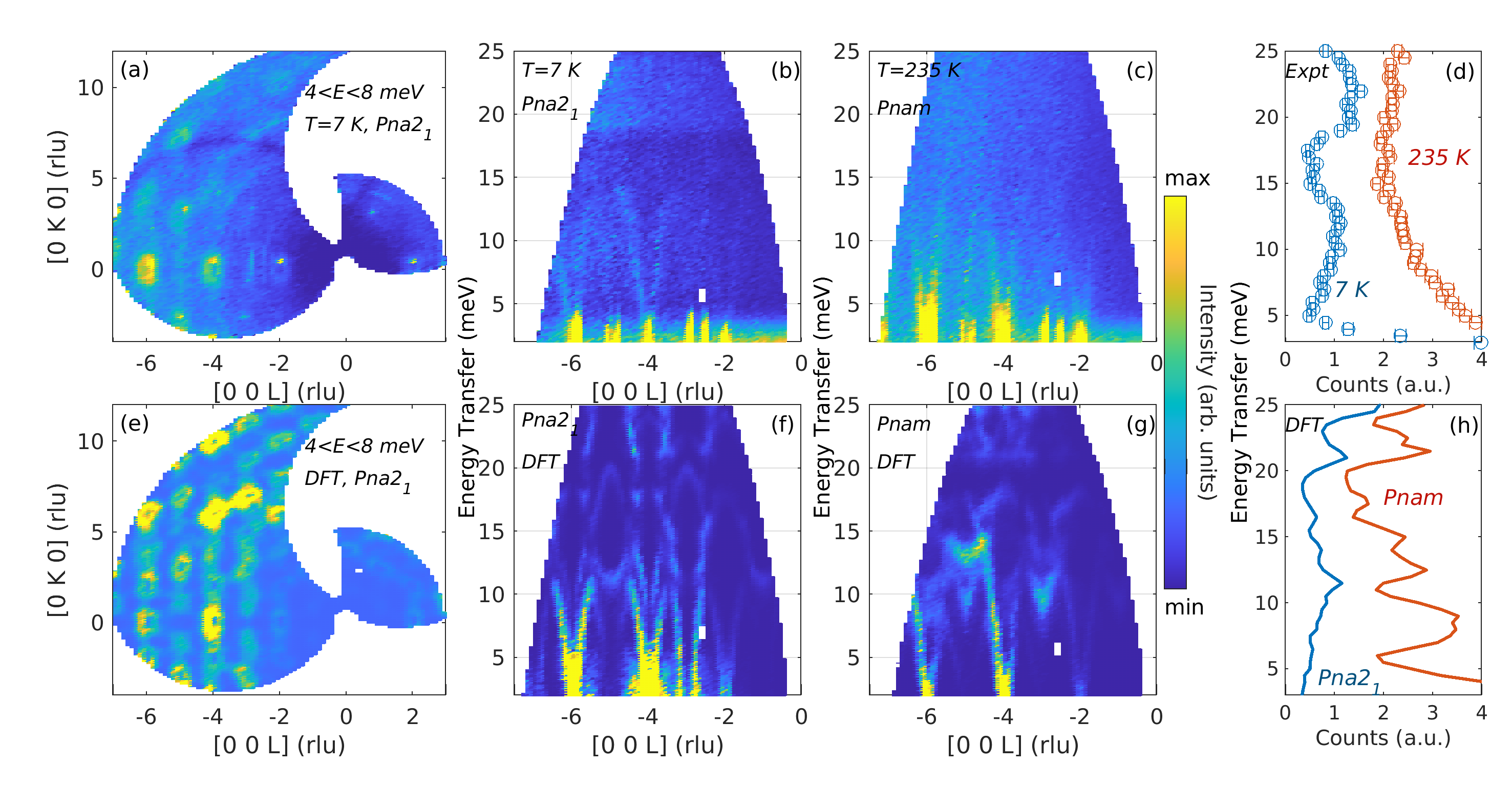}
\caption{A comparison of single-crystal phonon dispersion data (a--d) and simulations (e--f) for \amsulfd:
  (a, e) $(0KL)$ reciprocal space maps over $4<E<8$~meV, showing the emergence of cones of scattering from the acoustic phonons.
  (b, c, f, g) scattering intensity as a function of energy transfer along $[00L]$ in the (b, f) $Pna2_1$ and (c, g) $Pnam$ phases.
  (d, h) energy transfer cuts through $(0,0,3.5)$ in each phase. The color scale in the $L$-$E$ colormaps represents the intensity of the scattering. The simulation for the cut in $Pnam$ has been horizontally offset by 0.5 units along the $x$-axis for clarity (h), but no offset has been applied to the equivalent data set (d), which is genuinely more intense than seen for $Pna2_1$.}\label{figZ}
\end{figure*}

Moving away from the elastic line, acoustic phonons emerge as cones of scattering from the Bragg peaks, appearing as rings of intensity for energy transfers between $4$ and $8$~meV (Fig.~\ref{figZ}a). When the DFT results are weighted by the neutron scattering lengths of the respective nuclei and convolved with the instrumental resolution function, they agree reasonably well with the experimental data (Fig.~\ref{figZ}e). In particular, the simulations accurately reproduce the asymmetric shape of the rings of scattering, and hence the dispersion curves of the acoustic modes. For instance, the different orientations of the rings around $(0,10,-2)$, $(0,7,-5)$, $(0,3,-5)$, $(0,0,-4)$, and $(0,-3,-5)$ are well reproduced, as well as the regions of higher scattering intensity at the top of $(0,3,-5)$ and the top and bottom of $(0,-3,-5)$. 
Notable differences include first, that many of the rings associated with weaker diffraction peaks are not clearly visible above the noise in the experimental data. Second, the rings from several peaks on the edge of the observable region of reciprocal space, in particular $(0,6,-6)$ and $(0, 10, -4)$, are predicted to be more intense than seen in the data; we tentatively attribute this to masking of unreliable detectors at the edge of the array. Third, the calculations also predict an intense broad diagonal scattering peak in the region $5<K<10$, which is not apparent in the experimental data; this may partly be the result of absorption causing a dark ring cutting through the data set at this point. We suggest that all of these discrepancies in intensity at very low energy transfer reflect the intensities of the relevant Bragg peaks more than the dispersion relations. 
Further comparisons of data and simulation at different energy cuts are given in Fig.~S8. 

Next, we mapped the scattered intensity as a function of energy transfer along the $c^*$ axis in each phase.
In the low temperature phase data (Fig.~\ref{figZ}b), the acoustic modes are clear, emerging from the Bragg peaks on the elastic line at $(0\, 0\, L)$ for $L$ even. The scattering is more intense as $L$ increases, since the 1-phonon scattering intensity includes a factor of $|\mathbf{Q}|^2$. The bright signal at the base of the plot is bleeding coming from the elastic line due to the colour scale required to make the inelastic signal visible. The agreement between the data and simulation (Fig.~\ref{figZ}f) for the acoustic modes is excellent, and it is also reasonable for the optic modes. There is a notable discrepancy at the elastic line, where scattering from $(0,0,L)$ peaks with $L$ odd is observed despite the fact that these peaks are systematically absent in both $Pna2_1$ and $Pnam$. We attribute this to multiple scattering from the large mosaic sample; this effect will be most visible at the elastic line and is not expected to contaminate our data elsewhere.

The agreement can be seen more clearly from a one-dimensional cut through the data at the $Z$-point ($0$ $0$ $-3.5$) as a function of energy transfer. The blue trace in Fig.~\ref{figZ}d shows the tail of the elastic line at low energy transfer, then two broad peaks in the scattering at $12$~meV and $22$~meV, separated by a dip at $17$~meV. Comparing this with the equivalent cut through the simulated scattering (Fig.~\ref{figZ}h) including the instrumental resolution, the elastic line tail is absent, since only the inelastic scattering has been calculated, and the trace is more structured, but it too has a broad peak at around $11$~meV and a dip at around $18$~meV. The calculation has two peaks at $21$ and $25$~meV as opposed to the single broad peak at $22$~meV in the data, but some broadening of the data could mask individual peaks.

The inelastic signal at $T=235$~K in the $Pnam$ phase is much more intense (Fig.~\ref{figZ}c) due to the increased phonon population associated with the Bose-Einstein distribution, and also appears to be more diffuse, making it harder to identify phonon modes above the low energy acoustic modes. In addition, the multiphonon scattering that contributes to the background increases in intensity with increasing temperature. The equivalent simulation of the scattering for $Pnam$ including the instrumental resolution (Fig.~\ref{figZ}g) reveals considerably more structure than observed in the data. The steeply dispersing acoustic modes at the lowest energy transfers are reproduced, but the features above $\sim10$~meV cannot be resolved. One possibility is that, in the higher temperature phase, phonon-phonon interactions reduce the phonon lifetime, which broadens the spectral lineshape, smearing out the signal compared to the calculations performed in the harmonic approximation. However, the fact that well-defined phonons are observed at all is again evidence that there is no orientational disorder, which would lead to very significant broadening from over-damping. Moreover, it is clear that the gap in the scattering at around $18$~meV, seen in both the data and simulation in $Pna2_1$, is absent in both in $Pnam$. Again, some signal at systematically absent $(0,0,L)$ peaks ($L$ odd) is visible at the elastic line, attributable to multiple scattering.

Comparing the same one-dimensional cut as a function of energy transfer at the $Z$-point ($0$ $0$ $-3.5$) (Fig.~\ref{figZ}d), the data in $Pnam$ shows no clear features, just a slowly decreasing count rate with increasing energy transfer. Meanwhile, the calculation shows a series of peaks (Fig.~\ref{figZ}h). The calculated peak below $5$~meV would be obscured in the data below the elastic line tails, but the calculated peak at 8~meV may be reflected in the data. The broader features seen in the calculation are not visible in the data, but this is likely to be a consequence of the reduced phonon lifetime at higher temperatures.

Our data cover many Brillouin zones, such that the dispersion can be plotted along high symmetry directions other than the $z$-axis, but this shows the features most clearly. Slices along other high-symmetry directions 
compared with calculated phonon dispersion curves are shown in 
Figs.~S9 and S10.

\subsection{Powder inelastic neutron scattering under hydrostatic pressure}

Finally, INS measurements were performed on a powder \amsulfd\ sample in a clamp pressure cell. Because of powder averaging, this technique gives data only 
as a function of the modulus of the scattered momentum $|\mathbf{Q}|$. Clear bands of inelastic scattering intensity are seen at base temperature ($10$~K) and ambient pressure, separated by gaps at $5$ and $18$~meV (Fig.~S11), 
reminiscent of that seen at $Z$ in the single-crystal data (Fig.~\ref{figZ}d). The $18$~meV gap is also clearly visible in the generalised phonon density of states (gDOS) calculated according to Eq.~\ref{eqn1} (Fig.~\ref{figgdosmodes}a).


The DFT calculations allow the modes in each phase to be described in detail, consistent with but expanding previous assignments from Raman scattering \cite{torrie1972raman, raman1976}.
In the low-symmetry $Pna2_1$ phase (Table~S3), the phonons up to around $30$~meV involve whole-body motion, with translations of both ammonium and sulfate coupled with sulfate librations. There are 45 of these modes: 8 ammonium ions with 3 translational degrees of freedom, and 4 sulfate ions with six translational and librational degrees of freedom, minus the 3 zero-frequency modes corresponding to translation of the whole crystal. Because the primary motion is molecular translation, these appear in Fig.~\ref{figgdosmodes}b as modes with relatively low contributions from libration, bond stretching or angle bending.

The calculated phonons between 18 and 30~meV include several ammonium ``rattling'' modes. In our powder data, the localised nature corresponds to a vibrational energy that is independent of momentum transfer, so that these appear as flat bands of scattering in $S(Q,E)$, resulting in sharp features in the gDOS (Fig.~\ref{figgdosmodes}a) between 18 and 30~meV. While the general trend for phonon scattering intensity varies as $|\mathbf{Q}|^2$, collective motion will result in a departure from this simple relationship. That the measured intensity of these modes varies simply as $|\mathbf{Q}|^2$ (Fig.~S11) implies a lack of phase coherence, further evidence for identifying these as ``rattling'' modes \cite{Koza}.

The calculated modes between $34$ and $50$~meV are, with one exception, ammonium librations.\footnote{The exception occurs due to longitudinal optic/transverse optic splitting of the translational modes just below this frequency range, in which ammonium and sulfate ions move in opposite directions. The electric polarisation arising from this motion increases the frequency of the respective longitudinal branches in the limit as $\Gamma$ is approached from $X$, $Y$, and $Z$, so that along each of these branches, that single translational mode has a higher energy than the lowest-energy ammonium libration modes.} 
The librational modes group together into sets of four adjacent modes, all of which involve the same set of ammoniums rotating about the same axis. For example, the modes at $34.2, 34.3, 34.5$ and $34.7$~meV all involve the N2 ammonium rotating about the $c$-axis. The fact that all four have such similar energies means, in effect, that the four motions are almost independent. 

Above the ammonium librational modes, our $Pna2_1$ calculations identify three sets of internal modes for the tetrahedral sulfate ions:
\begin{center}
\begin{tabular}{m{0.8cm}m{1.8cm}m{3.8cm}}
  $\nu_2$ & 8 modes & $52.5-55.9$~meV,\\
  $\nu_4$ & 12 modes & $71.5-75.0$~meV,\\
  $\nu_1$ & 4 modes & $113.2-113.4$~meV,\\
\end{tabular}
\end{center}
in reasonable agreement with the free-ion values (56, 76 and 122 meV respectively), and labelled according to the convention used by Nakamoto \cite{Nakamoto} (Fig.~S12). The next 36 modes between 120 and 142~meV are SO$_4$ $\nu_3$ modes coupled with ND$_4$ $\nu_4$ modes, followed by 16 ND$_4$ $\nu_2$ modes, 8 ND$_4$ $\nu_1$ modes and 24 ND$_4$ $\nu_3$ modes. In these internal ND$_4$ modes, N1 and N2 are degenerate, unlike in the lower-energy librational modes (Table~S1).

Neither the ammonium ``rattling'' modes nor the internal modes of the ammonium and sulfate ions change substantially across the phase transition in our calculation (Table~S4).


\begin{figure}
\centering
\includegraphics[width=0.5\linewidth]{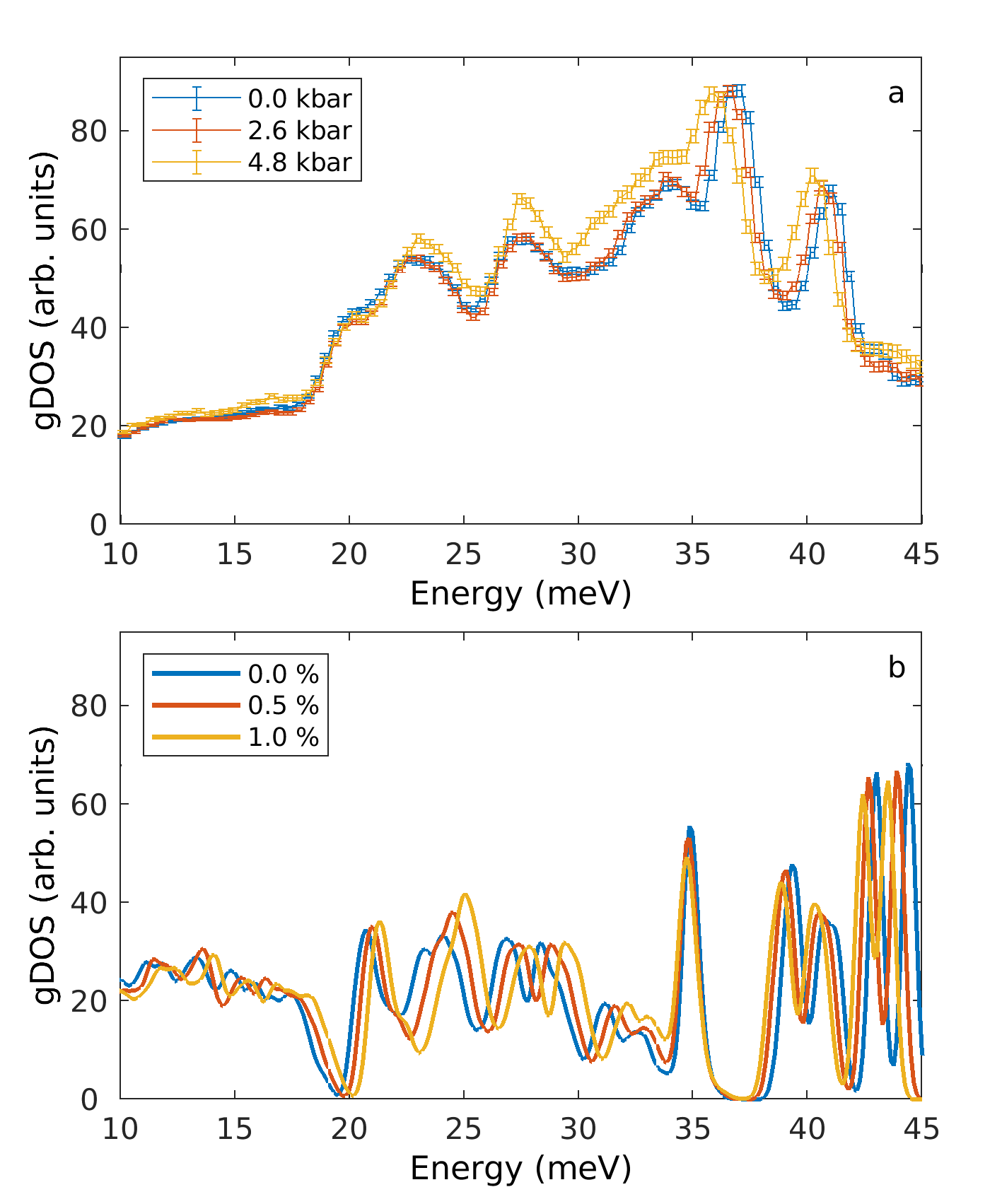}\\
\caption{(a) The neutron-weighted generalised phonon density of states of \amsulfd\ as a function of pressure, measured at $T=10$~K with $E_i=67$~meV, and (b) the DFT calculated phonon density of states in $Pna2_1$, as a function of strain. The low-frequency region shown corresponds to translations and librations of relatively rigid molecular ions. Most modes increase in frequency as the cell volume decreases (\emph{i.e.}, have positive mode Gr\"uneisen parameter); the ammonium librations at the upper end of the frequency range shown are easily distinguishable by their sharp features and decrease in frequency with decreasing volume (negative mode Gr\"uneisen parameter). In part (a), error bars represent the standard deviation $\sigma$ calculated using Mantid based on the Poisson statistics of discrete scattering events.}\label{figpressgdos}
\end{figure}

Due to finite phonon lifetimes, peaks in the density of states become increasingly broad and indistinct as the temperature increases (Fig.~S13), again consistent with previous Raman results \cite{raman1976}. To 
investigate the pressure dependence of the inelastic scattering, we therefore concentrated on low temperature data and hence the $Pna2_1$ phase. Figure~\ref{figpressgdos}a shows the generalised phonon density of states measured at base temperature ($T=10$~K) at $P=0.0, 2.6$ and $4.8$~kbar. This reveals that not all of the phonon modes exhibit the same pressure dependence. The majority of the modes at lower energies stiffen with pressure, as might be expected, since the atoms are brought closer together, increasing the forces between them. However the modes at $35-45$~meV soften with increasing pressure, indicating a negative mode Gr\"{u}neisen parameter, while the mode at $34$~meV shows no pressure dependence.


The same behavior is observed in our simulations for the $Pna2_1$ phase (Fig.~\ref{figpressgdos}b), with the whole body motion modes mostly stiffening, the polar translation mode at $33$~meV unaffected, and the ammonium libration modes having a negative Grüneisen parameter. 
Extending these calculations to different points in the Brillouin zone (Fig.~\ref{fig:dispersion}a) gives similar results. The more dispersive, lower-energy translation modes have either zero or positive mode Gr\"{u}neisen parameters, with the exception of the lowest-frequency branch along the path $\Gamma$--$X$--$S$: this involves transverse motion (\emph{i.e.}, in the $c$ direction) of the layers of molecules in the $ab$ plane, and has a mildly negative mode Gr\"uneisen parameter. On the other hand, the flatter ammonium librational modes between 38 and 45~meV show negative values of the mode Gr\"{u}neisen parameter. We rationalise the difference in these parameters by observing, analogous to the behaviour of framework solids predicted by the Rigid Unit Mode model \cite{Dove}, that the ammonium librations involve transverse motion with respect to the \ce{N\bond{-}H\bond{...}O} hydrogen bonds, and a reduced volume makes ``buckling'' more favourable, lowering the frequency. Negative mode Gr\"{u}neisen parameters may lead to negative thermal expansion, as is observed in ammonium sulfate just below the phase transition \cite{Shmytko}.

In $Pnam$ (Fig.~\ref{fig:dispersion}b), by contrast, almost all of the modes have positive Gr\"uneisen parameter. 
Although the eight potentially unstable modes for which we calculated effective harmonic frequencies at $\Gamma$ appear similar, they have rather different mode Gr\"uneisen parameters. For five of the eight modes, these are positive; for two, they are negative only at very low temperature but become positive thereafter; and for a single mode, the mode Gr\"uneisen parameter remains strongly negative ($\gamma = -4.7$) up to the experimental decomposition temperature of \SI{523}{K}. Thus, although modes involving libration of the same molecules about the same axis are similar in energy at ambient pressure, this result demonstrates that these modes couple very differently to strain, depending on the relative phase of the librations.


\section{Discussion and Conclusions}

Combining our experimental data and DFT model, we conclude first that there is no central barrier to the ammonium librations in the high-symmetry phase. Thus this phase is far from the classical disordered limit that would be accurately represented by a crystallographic split-site model. In particular, the Boltzmann entropy formula $S = R\ln n$ does not apply here, and we suggest that the similarity of the measured entropy to $(3\ln 2)R$ is ultimately a coincidence. This conclusion is in line with the warning given sixty years ago by Guthrie and McCullough, two pioneers of research on orientationally disordered crystals, that ``speculations based mostly on thermal data are [best] avoided in public''!\cite{guthrie_observations_1961}

Instead, the entropy is best understood as vibrational rather than configurational. Our results demonstrate that a flat-bottomed potential in which there is little resistance to small librations is consistent with the known structure and capable of producing an entropy of the correct magnitude. Such flat potential minima are reminiscent of the well shapes in the crystalline phases of inorganic semiconductor phase-change materials \cite{lencer_map_2008, matsunaga_phase-change_2011}. The structural origin of this dynamic entropy lies in the contrast between the relatively rigid hydrogen-bonded network in the low-symmetry phase -- as demonstrated by the harmonic librational frequencies in DFT, and by its experimental destabilisation under pressure -- and the more loosely held network in the high-symmetry phase that supports low-frequency librations. We note that, although one might loosely attribute these low frequencies to ``anharmonicity'', it is strictly speaking the small harmonic term in the potential expansion rather than the larger anharmonic terms that are responsible: indeed, a lower frequency, and higher entropy change, still would be possible if the quartic terms were \emph{lower}.

Second, we have determined in detail the phonon modes of ammonium sulfate.
There is excellent agreement between our experimental and simulation results in the low-temperature phase, even in the full four-dimensional $\mathbf{Q}$-$E$ space made available by the single-crystal data, suggesting that this methodology gives a reliable description of both internal and intermolecular contributions to the crystal energy. In the high-temperature phase, low phonon lifetimes obscure the experimental data and comparison becomes more difficult, but the agreement still appears to be reasonable. 
The good agreement at low temperatures suggests that the DFT simulations are a reliable guide to the phonon dispersion, even at temperatures where this is unclear from the experimental data.
%
The frequency of the ammonium librational modes is overestimated by DFT; these could nonetheless be conclusively identified in experiment by their negative Gr\"{u}neisen parameters. The frequency of the internal ammonium and sulfate modes is slightly underestimated but, again, identification is clear because these are isolated in frequency. 

Our results suggest a structural mechanism for the distinct change in the phonon spectrum between the $Pna2_1$ and $Pnam$ phases of ammonium sulfate that is responsible for the high entropy of the $Pnam$ phase and hence for the caloric effects in this material. In the low-entropy $Pna2_1$ phase, the ammonium librational modes from both experiment and simulation form distinct peaks, separate from the other low-frequency collective modes, indicating that hydrogen bonding holds these ions relatively firmly in place. Strong hydrogen bonding also means that the librational ammonium motion is able to drag neighbouring sulfate ions along with it (consistent with a proposed mechanism for the phase transition \cite{Malec}), causing the lattice to contract with temperature. This mechanism therefore also explains the negative mode Gr\"uneisen parameters that are, again, observed in both experiment and simulation; similarly, it suggests that these modes are also likely to be responsible for the region of negative thermal expansion just below the phase transition. By contrast, in the high-entropy phase where the ammonium ions sit on the newly created mirror plane, they are less firmly held in place, leading to lower vibrational frequencies that exert less influence over the shape and size of the lattice.

In this work we have not considered the contribution of the phonons to the thermal conductivity, but this too will be important to practical applications, reinforcing the importance of studying the phonon behaviour under working pressure conditions.


Ammonium sulfate is a particularly clear example of the perils of ignoring dynamic contributions to entropy, since, as we have shown, the contribution from configurational disorder appears to be negligible. However, these results are also relevant to other materials where configurational contributions are also important. Many materials that undergo high-entropy phase transitions -- including the globular organic ``plastic'' crystals and the molecular perovskites -- have an orientationally ordered state held together by weak interactions such as hydrogen bonds and a disordered state in which these bonds are broken. While it is convenient for the purposes of entropy calculations to consider ``pure'' order-disorder transitions, the distinction between these and displacive phase transitions is in reality a continuum \cite{bruce_cowley_structural}. It thus seems likely that the Boltzmann formula gives a misleading picture of the true origins of entropy in many or even most molecular crystals. The example of ammonium sulfate further suggests a different paradigm for crystal engineers to target in the search for high-entropy and caloric materials: one where \emph{competing} networks of hydrogen bonds or other weak interactions instead create a complex, high-entropy energy landscape.


\section*{Methods}
\subsection*{Calculation}
\label{sec:dft}
We used density functional theory to calculate the structures and phonon dispersion relations of both phases of ammonium sulfate, as implemented in the CASTEP software package, v.19.11 academic release \cite{Castep_Clark,Castep_Refson}. We used a plane-wave basis set together with norm-conserving pseudopotentials from the CASTEP standard library. The Perdew-Burke-Ernzerhof functional was used to describe the exchange-correlation energy \cite{PBE}, with a Tkatchenko-Scheffler empirical dispersion correction to account for the van der Waals interaction between the molecular ions \cite{SEDC-TS}. 

In the geometry optimization, the energy cutoff was set to $1100$~eV, and a $2\times1\times2$ Monkhorst-Pack $k$-point grid used. The structures were relaxed until both the forces between atoms became smaller than $0.01$~eV/A and the energy change between steps was less than $2\times 10^{-5}$~eV/atom. 
%
%
For the phonon calculation, the force matrix was calculated by density-functional perturbation theory, using a finer $3\times3\times3$ phonon $k$-point grid. As discussed in the Results section below, in the $Pnam$ phase using experimental lattice parameters (although not when using relaxed lattice parameters, nor in the $Pna2_1$ phase) this gave several unstable modes with $\omega^2 < 0$ in the harmonic approximation. To take these modes into account, we mapped the energy $V(\eta)$ as a function of the normal coordinate $\eta$ of each mode individually at $\Gamma$, then fitted the results to a polynomial potential model. 
To find the energy eigenstates of this model, we followed the approach of Skelton and co-workers \cite{skelton_anharmonicity_2016}, using the Fourier grid Hamiltonian method \cite{marston_fourier_1989} to solve the relevant Schr\"odinger equation \cite
{bruesch_phonons_1982} numerically: 
\begin{equation}
    -\frac{\hbar^2}{2}\frac{\mathrm d^2 \psi(\eta)}{\mathrm d \eta^2} + V(\eta)\psi(\eta) = E\psi(\eta).
\end{equation}
From these eigenstates, we calculated the temperature-dependent partition function
\begin{equation}
    Z(T) = \sum_i\exp\left(-\frac{E_i}{kT}\right)
\end{equation}
and hence an effective harmonic frequency, which is therefore also now a function of temperature \cite[p. 226]{bruesch_phonons_1982}: 
\begin{equation}
    Z = \left[\exp\left(\frac{\hbar\omega}{2kT}\right)
    -\exp\left(-\frac{\hbar\omega}{2kT}\right)
    \right]^{-1}
\end{equation}
\begin{equation}
    \omega_\text{eff}(T) = \frac{2kT}{\hbar}\sinh^{-1}\left(\frac{1}{2Z}\right)
\end{equation}
For simplicity, and because following a mode in this way at an arbitrary point of the Brillouin zone requires a costly supercell calculation, we assumed that these effective frequencies were constant with wavevector. Using a $2\times1\times1$ supercell to investigate the $X$ point suggested that incorporating more points in reciprocal space has a negligible effect within the error of our model, and does not change our qualitative conclusions.

To compare the DFT calculations and the experimental INS data, we used Euphonic \cite{Euphonic}, a Python package that efficiently calculates phonon bandstructures and inelastic neutron scattering intensities from a force constants matrix using the 1-phonon scattering function, such that one can obtain neutron weighted phonon dispersions and the density of states from precalculated phonon frequencies. This can have a stark effect on the range of visible modes in the simulated phonon dispersions, both due to the variation in coherent scattering cross-sections for different elements, and the polarisation factor $\mathbf{Q\cdot e}$ in the 1-phonon coherent scattering cross-section, where $\mathbf{e}$ is the phonon mode eigenvector. We also used TobyFit within Euphonic to include the experimental resolution function in our simulations.

The mode Gr\"uneisen parameters were calculated according to the standard equation
\begin{equation}\label{eq:gru}
    \gamma_i = -\frac{\partial\ln\omega_i}{\partial\ln V}.
\end{equation}
For the phonon calculations performed using experimental rather than optimized cell parameters, we were therefore unable to evaluate this expression under hydrostatic pressure; instead, we applied uniform strains of $\pm0.2\%$ in each dimension, using the central difference algorithm to numerically evaluate the derivative in (\ref{eq:gru}). Using uniform strain instead of hydrostatic pressure is equivalent to assuming that the material is elastically isotropic, which -- especially at the qualitative level of our analysis here -- is comparable to the other approximations involved in this calculation.

\subsection*{Inelastic neutron scattering}
Ammonium sulfate of natural isotopic abundance was purchased from Sigma Aldrich and deuterated by four successive recrystallisations from \ce{D2O}. Assuming uniform mixing of hydrogen atoms, this should give a $97.7\%$ deuterated sample. In practice, the incoherent background from $^1$H was not observed to any significant extent in our neutron scattering data. Single crystals with dimensions $\sim5\times30\times1$~mm$^3$ were obtained by precipitation from a saturated deuterated aqueous solution of \amsulfd.

Two sets of inelastic neutron scattering experiments were performed on fully deuterated ammonium sulfate using the Merlin spectrometer at ISIS \cite{Bewley}. First, single crystal measurements 
were performed at ambient pressure using an array of crystals of total mass $1.2$~g. These were coaligned by crystal habit, confirmed using the ALF alignment facility at ISIS, and attached using cytop to two aluminium plates so that sample covered an area of $30$ mm $\times$ 40 mm, oriented with a horizontal scattering plane defined by the $b$ and $c$ crystallographic axes and a mosaic spread of less than $5$~degrees (Fig.~S6). The plates were then stacked and inserted into the CCR.  

In the second set of experiments, a $1.3$~g sample of polycrystalline \amsulfd\ was loaded into a newly commissioned, low-background TAV6 cylindrical clamp pressure cell \cite{Kibble}, using helium as the pressure transmitting medium.
This cell has a cylindrical geometry, with walls of 5.3 mm and inner bore with a 7.0 mm diameter. The cell was inserted into a cylindrical shaped radial collimator coated with gadolinium paint, to minimise the background scattering from the pressure cell. This in turn was mounted in a closed cycle refrigerator, allowing data to be collected in the ranges $T=10$--280~K and $P=0$--4.8~kbar.

Merlin was operated in multi-rep mode, with incident neutron energies of $23$, $36, 67$ and $162$~meV. The raw data were processed using Mantid \cite{Arnold}, following the standard conventions \cite{Windsor}. For the single-crystal measurements, the sample was measured in discrete angular steps ($0.5^\circ$) about the vertical axis [$H$00] to form a ``Horace-scan'' over $90$~degrees in total, enabling a large region of energy-reciprocal space to be explored. The processed single crystal data was then combined using Horace \cite{Horace} to create $S(\mathbf{Q},\omega)$. Attempts to subtract off the background scattering arising from the aluminium plates proved unsuccessful, but there is little evidence of inelastic scattering from the aluminium in our data, just streaks of elastic scattering which can easily be differentiated from the Bragg peaks from the sample.


For the polycrystalline measurements, background data were collected by measuring the TAV6 cell filled with helium, under the same pressure and temperature conditions.\footnote{It is essential to include the helium in the background measurements, since it has a strong phonon signal in the solid state, and a weaker signal in the liquid state.} This was then subtracted from the relevant data sets. The powder data were also corrected for the absorption from the TAV6 cell to give the inelastic scattering function $S(Q,\omega)$.

The phonon density of states is calculated in Mantid using the 1-phonon scattering function formula in the incoherent approximation \cite{Chaplot}:
\begin{equation}\label{eqn1}
  S_{inc}(Q,E)=\exp(-2\bar{W}(Q))\frac{Q^2}{E}\left\langle n+\frac{1}{2}\pm\frac{1}{2}\right\rangle 
              \cdot\left[\sum_j\frac{\sigma^\text{scatt}_j}{2m_j}g_j(E)\right],
\end{equation}
where the term in square brackets is the calculated-neutron weighted density of states, and $g_j(E)$ is the partial density of states for each element $j$ in the material, and $m_j$ is the relative atomic mass of the component. The average Debye-Waller factor $\exp(-2\bar{W}(Q))$ is calculated using an average mean-square displacement $\langle u^2\rangle$, using
\begin{equation}\label{eqn2}
  W=\frac{Q^2\langle u\rangle^2}{2}.
\end{equation}

\subsection*{Total scattering and reverse Monte Carlo analysis}
Neutron total scattering data were measured from a perdeuterated powder sample of ammonium sulfate on the Polaris diffractometer at the ISIS Neutron and Muon Source, U.K. The sample was loaded in a vanadium can of \SI{6}{mm} diameter within a CCR. A series of 15-minute short runs were measured to extract the lattice parameters, and 6-hour-long runs were measured at six temperatures (150 K, 180~K, 210 K, 240 K, 270 K and 300 K) for total scattering data measurements. The data were focused using Mantid \cite{Arnold}.

The Reverse Monte Carlo simulation was performed using the RMCProfile code \cite{tucker_rmcprofile_2007}. Configurations of each phase used a $6\times 5\times 8$ supercell containing \num{14400} atoms, initially arranged according to the average crystal structure from Rietveld refinement. The RMC simulations were carried out for at least \num{5e6} steps in total (at least \num{1.2e6} accepted moves) until convergence. To prevent ions from distorting to physically unreasonable shapes, internal molecular bond and angle potentials, taken from the MM3 parametrisation \cite{MM3}, were used. To avoid unphysical spikes in the partial PDFs, a soft ``curvature'' restraint was placed on the second derivative.

\section*{Data availability}
Raw data from the neutron scattering experiments are available at \url{https://doi.org/10.5286/ISIS.E.RB1820305}, \url{https://doi.org/10.5286/ISIS.E.RB1910408}, \url{https://doi.org/10.5286/ISIS.E.RB1910572}, \url{https://doi.org/10.5286/ISIS.E.RB1920740}, and \url{https://doi.org/10.5286/ISIS.E.RB2000267}.

\section*{Acknowledgements}
The authors are indebted to Rebecca Fair (STFC) and M. Duc Le (ISIS) for their assistance and guidance in the use of Euphonic; to Chris Goodway and Mark Kibble (ISIS) for supporting the high pressure experiments; to Jeremy K. Cockcroft (University College London) for diffraction measurements to help co-align the crystals for the INS measurements; and to Jonathan M. Skelton (Manchester) for helpful discussions about the DFT calculations. We thank ISIS Neutron and Muon Source for the award of beamtime (RB1820305 and RB1910572 on MERLIN; RB1910408 on POLARIS; RB1920740 and RB2000267 on OSIRIS). AEP, HCW and BEM thank ISIS Neutron and Muon Source and Queen Mary University of London for funding a Facilities Development Studentship. SY, GC, AEP and MTD thank the China Scholarship Council for studentship funding. AEP, HCW and RJCD acknowledge EPSRC for funding (EP/S03577X/1). For computational resources, we are grateful to the UK Materials and Molecular Modelling Hub, which is also partially funded by EPSRC (EP/P020194/1 and EP/T022213/1).


\bibliographystyle{angew}

\providecommand*{\mcitethebibliography}{\thebibliography}
\csname @ifundefined\endcsname{endmcitethebibliography}
{\let\endmcitethebibliography\endthebibliography}{}

\newpage

\begin{minipage}[t]{55mm}
\vspace{0pt}
\includegraphics[width=55mm]{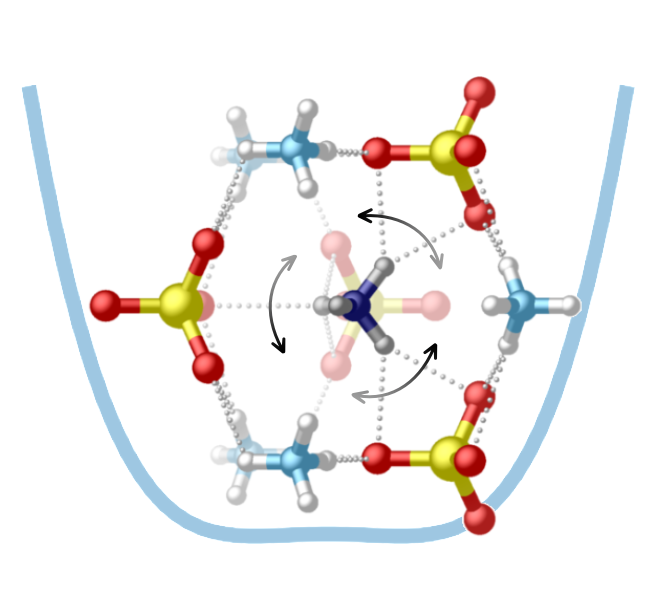}
\end{minipage}
\hspace{0.5em}
\begin{minipage}[t]{10.5cm}
\vspace{36pt}\sffamily
The large entropy change in ammonium sulfate, historically analysed in terms of an order-disorder phase transition, is shown instead to arise from low-frequency librations of ammonium molecules in flat-bottomed energy wells. This suggests that similarly competing hydrogen-bonded networks may be an attractive target for engineering new caloric molecular-ionic materials.
\end{minipage}

\end{document}


\begin{center}
  \textsf{\Large \textbf{Origin of the large entropy change in the molecular caloric and ferroelectric ammonium sulfate
}}
  
  \vspace{1ex}

  \textsf{
    Shurong Yuan,\footnotemark[2] 
    Bernet E. Meijer,\footnotemark[2]
    Guanqun Cai,\footnotemark[2]
    Richard J. C. Dixey,
    Franz Demmel,
    \mbox{Martin T. Dove},
    Jiaxun Liu,
    Helen Y. Playford,
    Helen C. Walker,\footnotemark[1] and
    Anthony E. Phillips\footnotemark[1]}

  \vspace{2ex}

  \textsf{\textbf{\large Electronic Supplementary Information}}
\end{center}

\tableofcontents

\footnotetext[1]{a.e.phillips@qmul.ac.uk; helen.c.walker@stfc.ac.uk}
\footnotetext[2]{These three authors contributed equally as joint first co-authors.}

\newpage

\section{Gr\"uneisen parameters in $Pnam$}
\begin{figure}[h]
    \centering
    \includegraphics
    {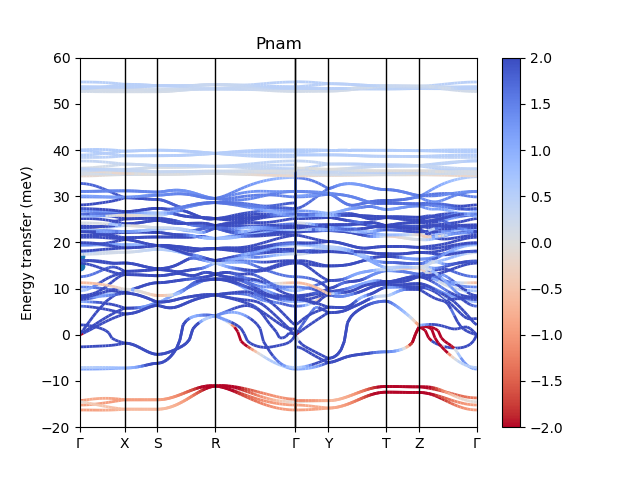}
    \caption{The phonon dispersion curve from DFT in the $Pnam$ phase, calculated in the harmonic approximation and coloured to indicate the mode Gr\"uneisen parameter at each point.}
    \label{fig:Pnam-gru-full}
\end{figure}

As discussed in the main text, in the $Pnam$ phase, eight modes are unstable in at least some parts of the Brillouin zone. Figure~\ref{fig:Pnam-gru-full} shows the dispersion curve in this phase \emph{without} correcting for this in any way; thus the unstable modes are shown with negative frequency, and the ``mode Gr\"uneisen parameters'' indicated by the colour scale are in fact calculated only from the variation of the harmonic term.

\newpage

\section{RMC models}
\begin{figure}[h]
    \centering
    \includegraphics[width=0.48\textwidth]{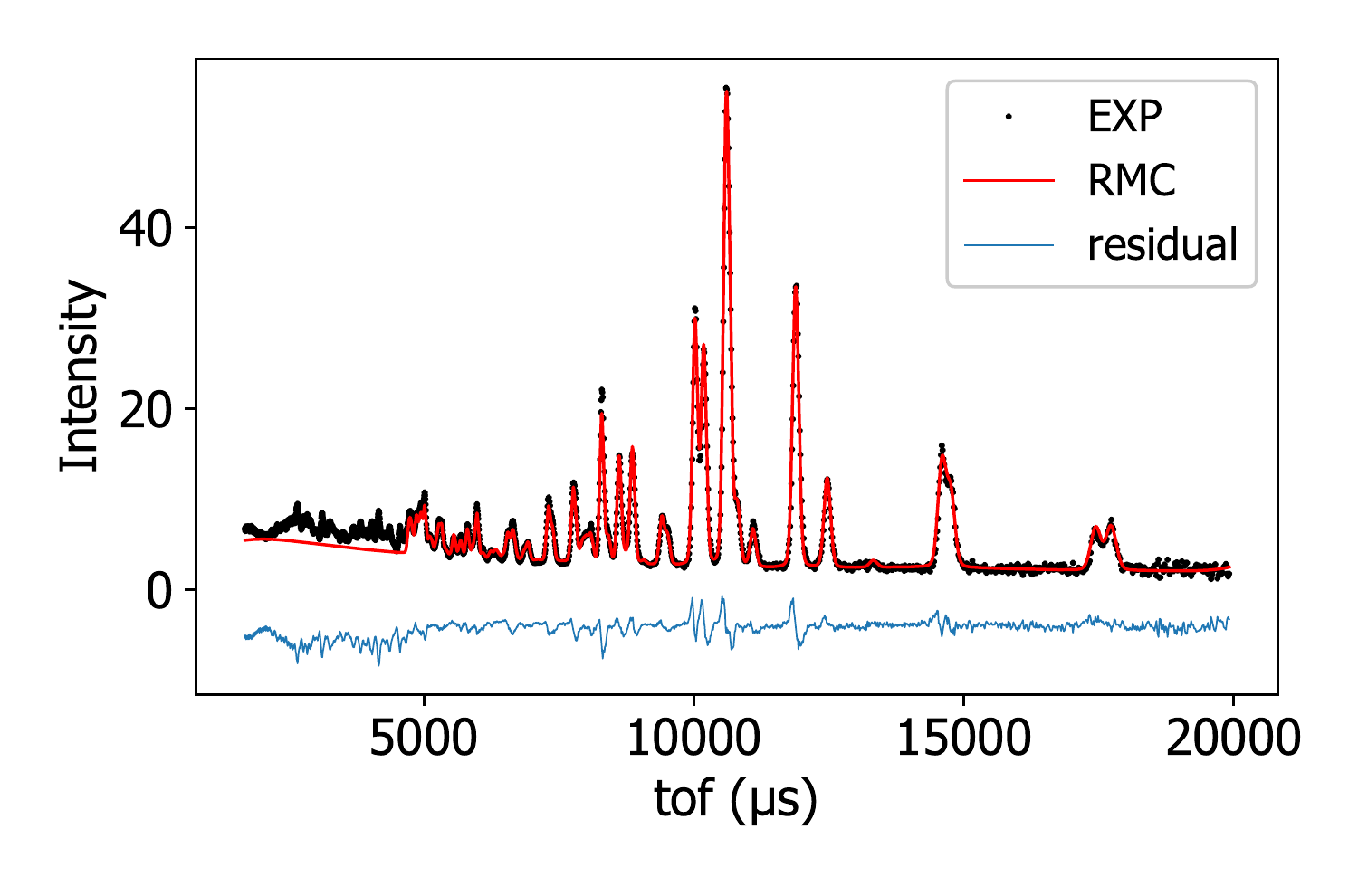}\\
    \includegraphics[width=0.48\textwidth]{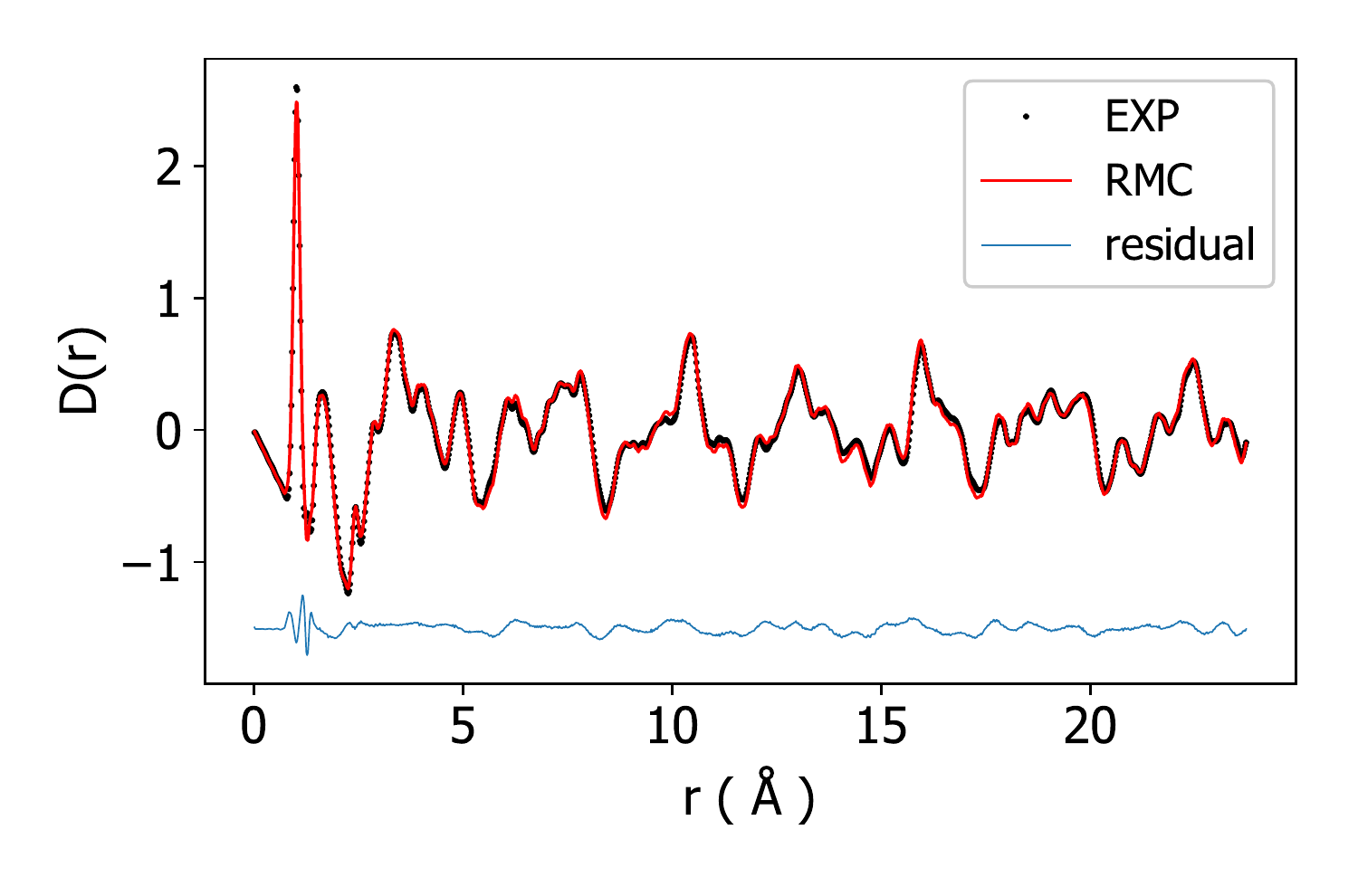}\\
    \includegraphics[width=0.48\textwidth]{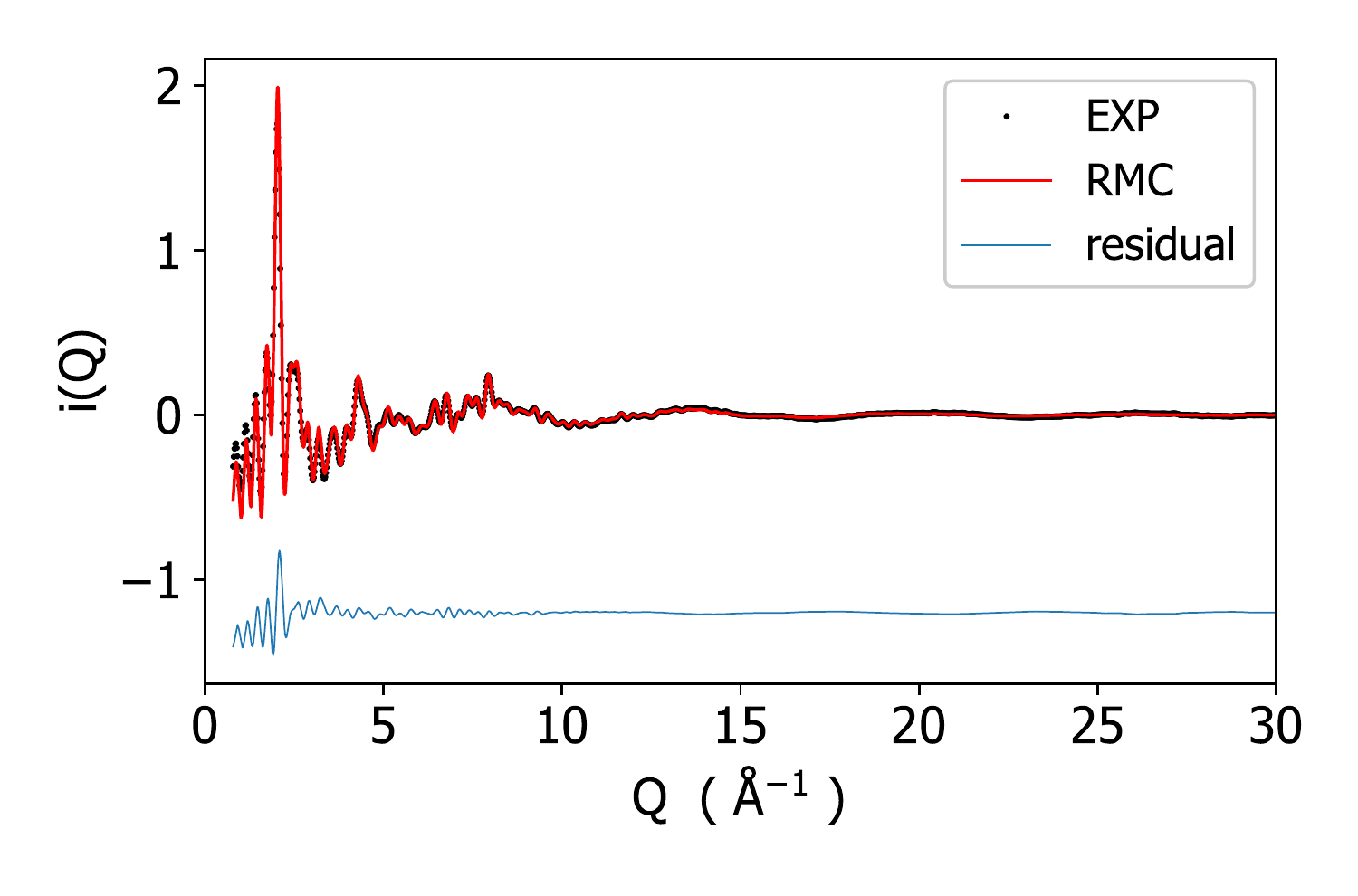}
    \caption{Representative fits from simultaneous reverse Monte Carlo modelling against three data sets: (top) the Bragg profile; (middle) the pair distribution function $D(r)$; and (bottom) the scattering function $i(Q)$. Shown here are the 150~K data sets.}
    \label{fig:RMC-fit}
\end{figure}

Representative fits from reverse Monte Carlo modelling are shown in Figure~\ref{fig:RMC-fit}.

\newpage

\section{Motion in the unstable modes}

\begin{figure}[h]
    \centering
        \begin{tabular}{ccc}
             \includegraphics[width=3.2cm]{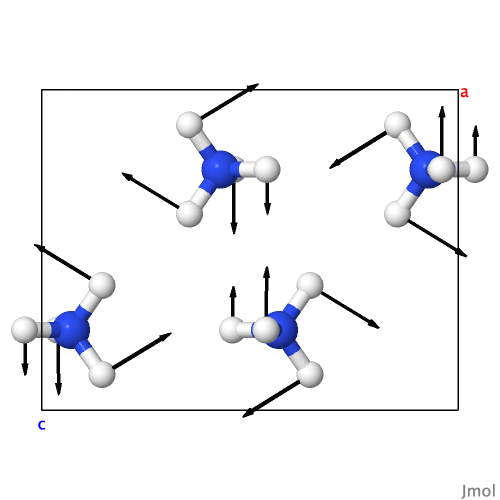} &
             \includegraphics[width=3.2cm]{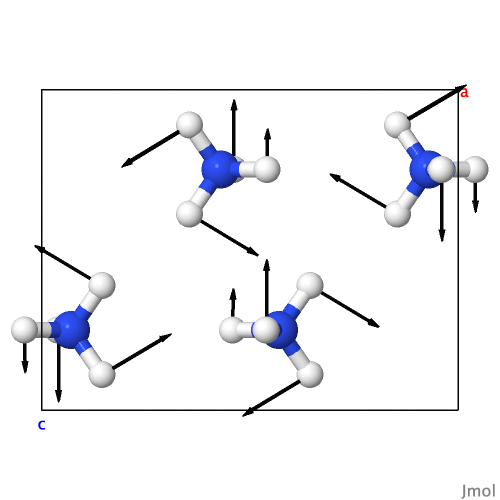} \\
             \includegraphics[width=3.2cm]{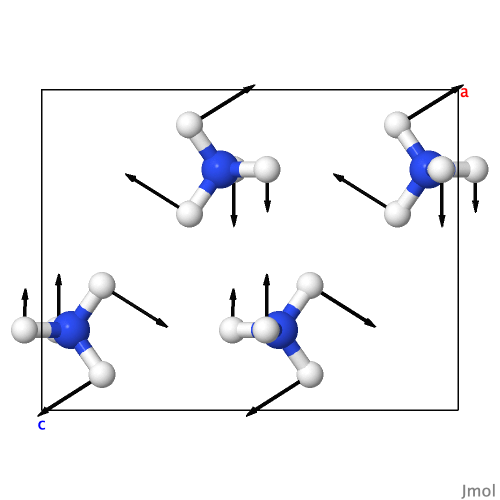} &
             \includegraphics[width=3.2cm]{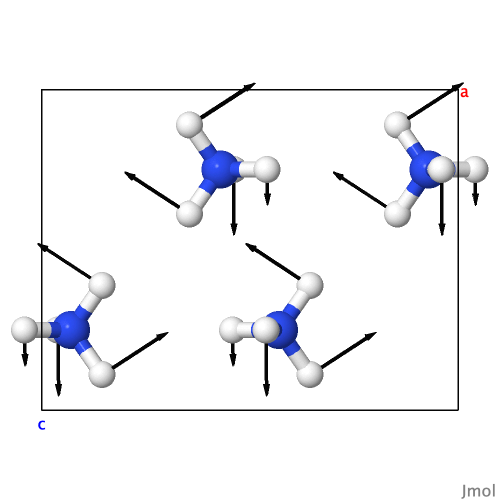} &
             \includegraphics[width=3.2cm]{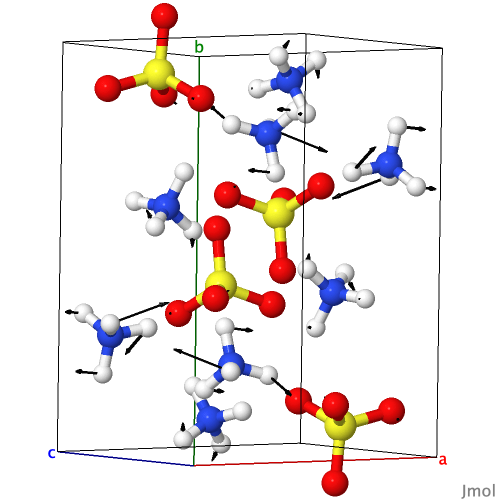} \\ \midrule
             \includegraphics[width=3.2cm]{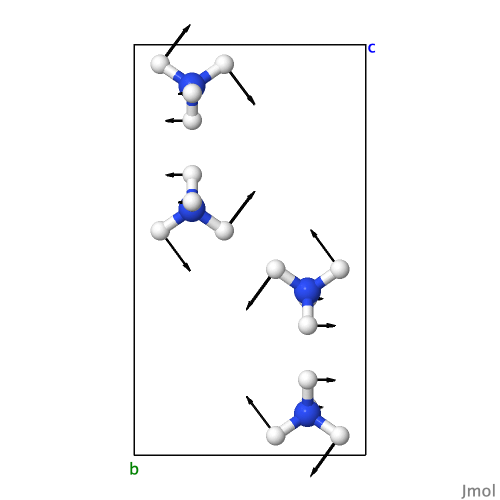} &
             \includegraphics[width=3.2cm]{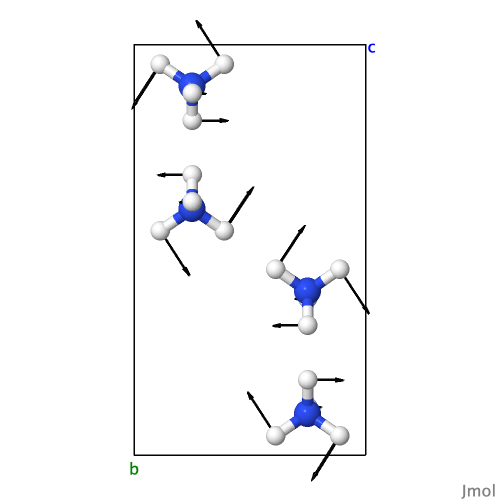} \\
             \includegraphics[width=3.2cm]{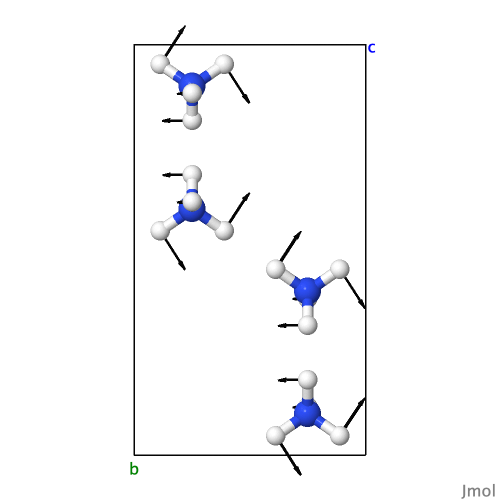} &
             \includegraphics[width=3.2cm]{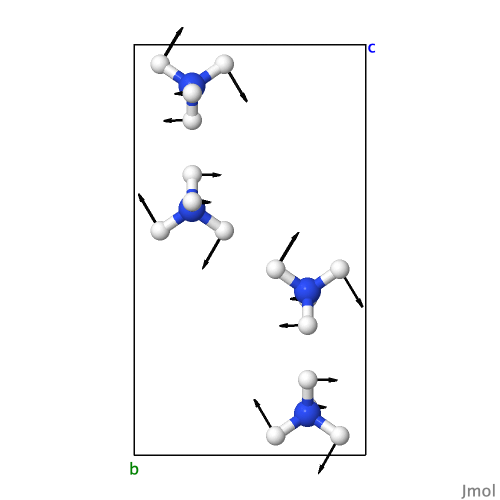} & 
             \includegraphics[width=3.2cm]{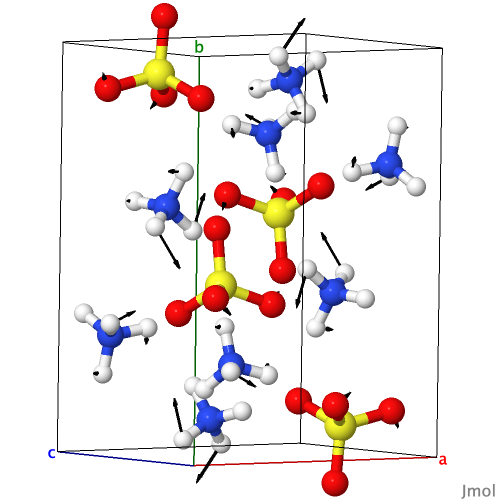} \\
        \end{tabular}
    \caption{Illustrations of the eight unstable modes. Top: \ce{N(2)H4+} librations about $b$; bottom: \ce{N(1)H4+} librations about $a$. In both cases the diagrams on the left show the four modes of this sort, with only the ammonium ions that move the most drawn; the diagram on the right shows the first mode of the set with all atoms drawn, showing the much smaller motion of the remaining atoms.}
    \label{fig:mode-diagrams}
\end{figure}

The \ce{NH4+} librational modes that become unstable in the harmonic approximation, and that are therefore responsible for the bulk of the entropy change, are illustrated in Fig.~\ref{fig:mode-diagrams}. 

\newpage

\section{QENS models}

\begin{figure}[h]
    \centering
    \includegraphics[width=0.5\textwidth]{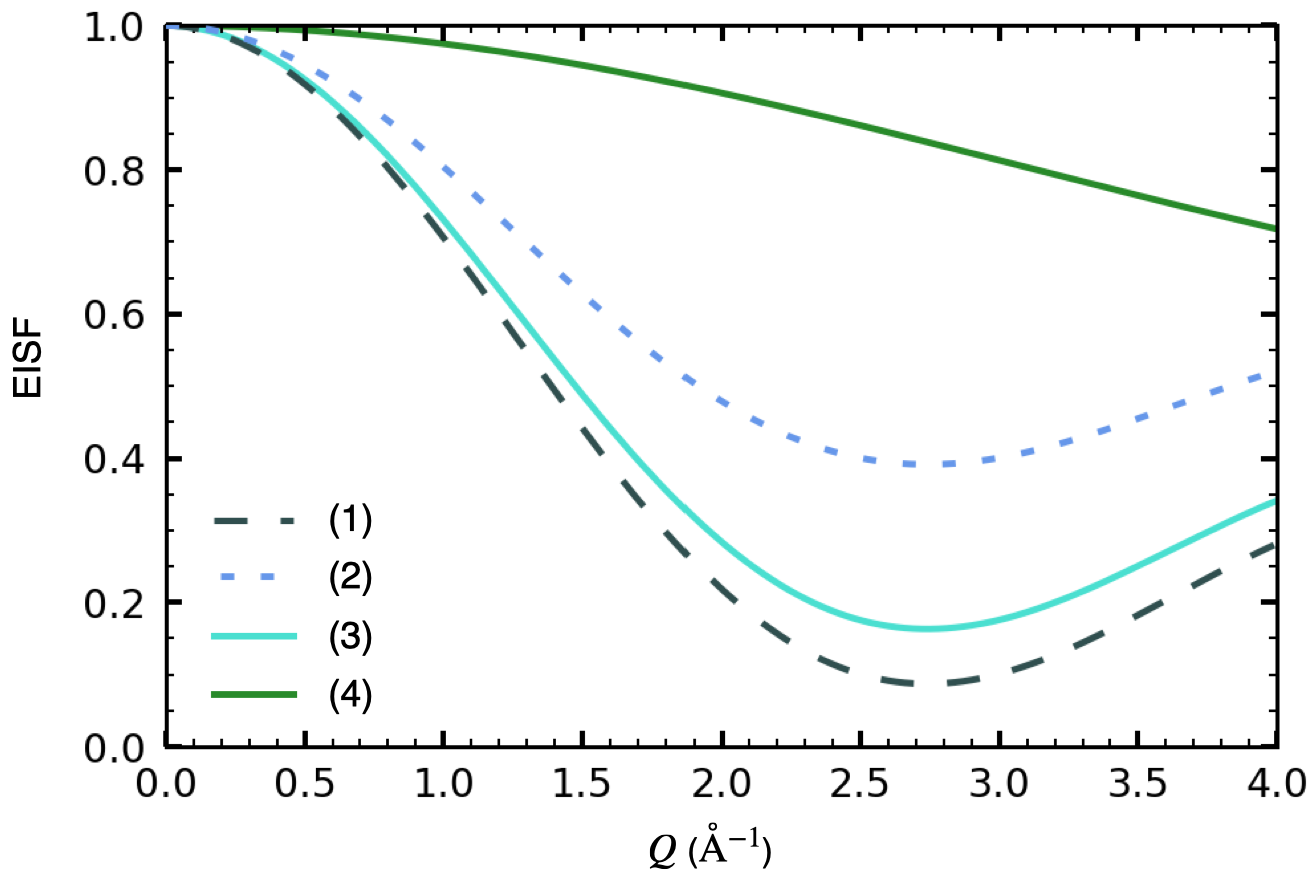}
    \caption{EISF models for the reorientational geometry with $f=m=1$. The models are (1) tetrahedral, (2) C\textsubscript{2}/C\textsubscript{3}, (3) Goyal 1978 and (4) mirror plane hopping.}
    \label{fig:eisfmodels}
\end{figure}

The models shown in figure \ref{fig:eisfmodels} are fitted to the experimental EISFs, with $f$ and $m$ as the free parameters. The fitted parameters are given in tables~\ref{tab:QENS200K} and~\ref{tab:QENS300K}.

\begin{table}[h]
  \centering\footnotesize
  \begin{tabular}{c|c|c|}
     Model & $f$ & $m$ \\
     \hline
     (1) tetrahedral & 0.49 $\pm$ 0.01 & 0.47 $\pm$ 0.03 \\
     (2) C\textsubscript{2}/C\textsubscript{3} & 0.61 $\pm$ 0.02 & 0.57 $\pm$ 0.02 \\
     (3) Goyal 1978 & 0.51 $\pm$ 0.01 & 0.49 $\pm$ 0.025 \\
     (4) mirror plane & 1 $\pm$ 0.001 & 0.67 $\pm$  0.08 \\
\end{tabular}

\caption{200 K}
\label{tab:QENS200K}
\end{table}

\begin{table}[h]
  \centering\footnotesize
  \begin{tabular}{c|c|c|}
     Model & $f$ & $m$ \\
     \hline
     (1) tetrahedral & 0.9 $\pm$ 0.01 & 0.84 $\pm$ 0.02 \\
     (2) C\textsubscript{2}/C\textsubscript{3} & 1 $\pm$ 0.1 & 0.75 $\pm$ 0.05 \\
     (3) Goyal 1978 & 0.97 $\pm$ 0.01 & 0.85 $\pm$ 0.01 \\
     (4) mirror plane & 1 $\pm$ 0.3 & 0.53 $\pm$  0.3 \\
\end{tabular}

\caption{300 K}
\label{tab:QENS300K}
\end{table}

\begin{figure}[h]
    \centering
    \includegraphics[width=\linewidth]{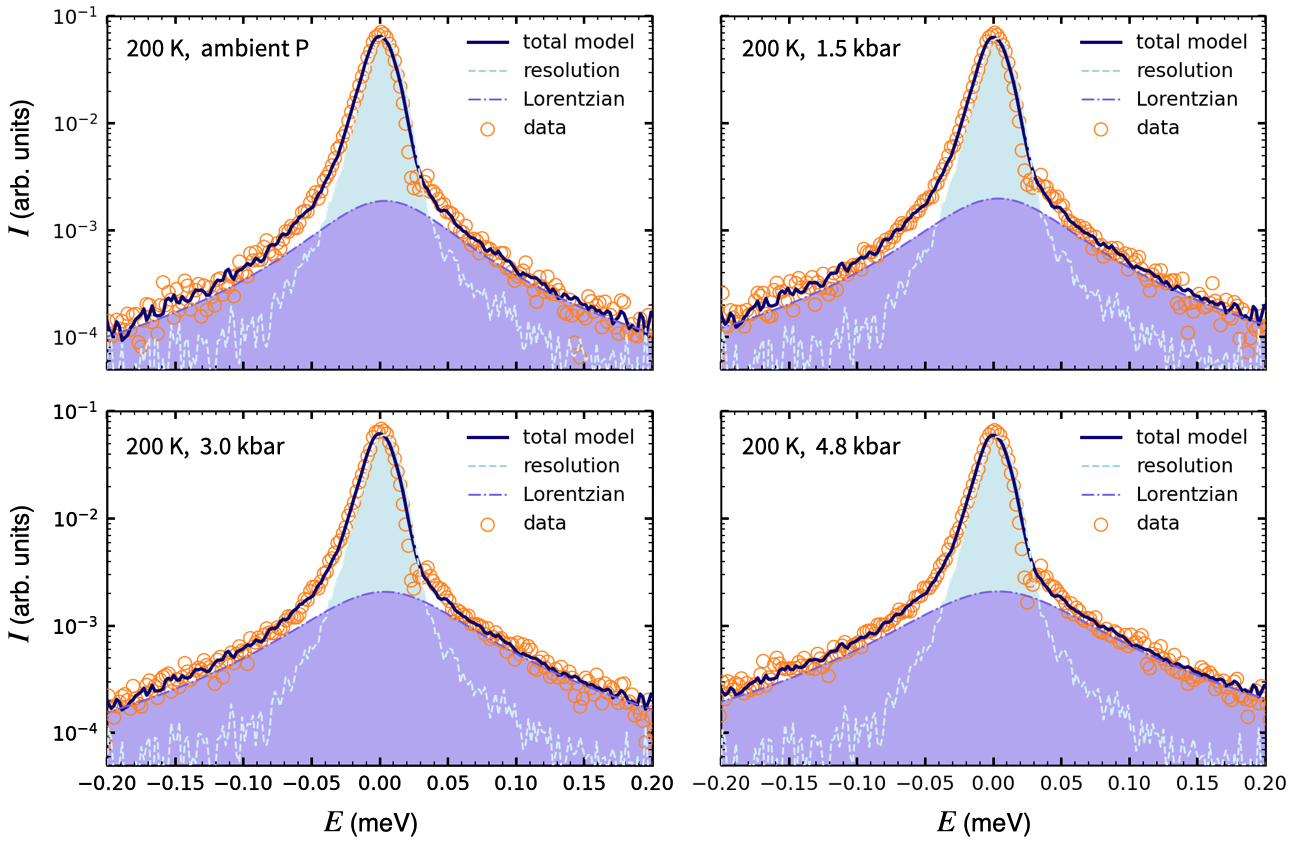}
    \caption{Fits of the structure factor (eq. 9, main text) to the experimental spectra (summed over $0 < Q < \SI{2}{\angstrom^{-1}}$) in the $Pna2_1$ phase.}
    \label{fig:SI_qens}
\end{figure}

\newpage
The fits to the pressure cell data in the $Pna2_1$ phase are shown in figure \ref{fig:SI_qens}.

\newpage

\section{Single crystal \amsulfd}
\subsection{Sample Alignment}
\begin{figure}[bhp]
    \centering
    \includegraphics{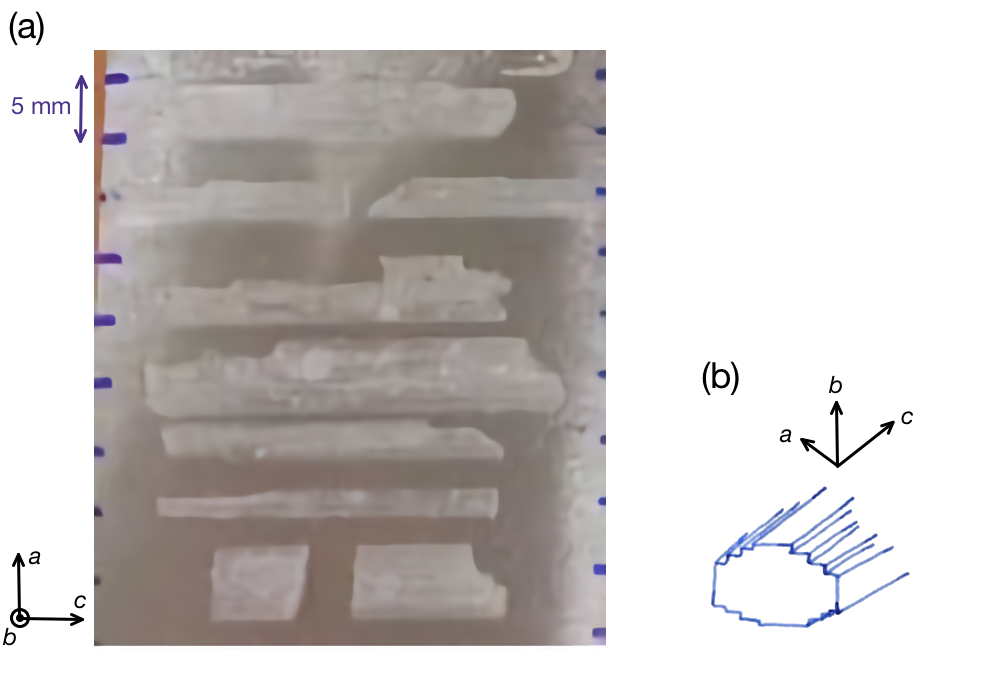}
    \caption{(a) Annotated photograph of one of the two aluminium plates, with deuterated ammonium sulfate crystals attached using cytop, used for our single crystal inelastic neutron scattering measurements. (b) Sketch of the single crystal habit.}
    \label{fig:photo}
\end{figure}

The habit of the deuterated single crystals as grown facilitates their co-alignment (see Fig.~\ref{fig:photo} (b)). This was further checked using single crystal neutron diffraction on the ALF instrument at ISIS. While the crystal axes are very clearly defined, the crystal habit, without a large flat facet, made it difficult to avoid introducing a tilt about the $c$ axis on mounting the crystals. The crystals were co-aligned to give a $(0KL)$ horizontal scattering plane, and the crystal mosaic was minimised as far as possible to give the smallest spots on the detectors associated with the Bragg reflections.

\newpage

\subsection{INS Experiment}
\begin{figure}[b]
\centering
\includegraphics[width=0.65\linewidth]{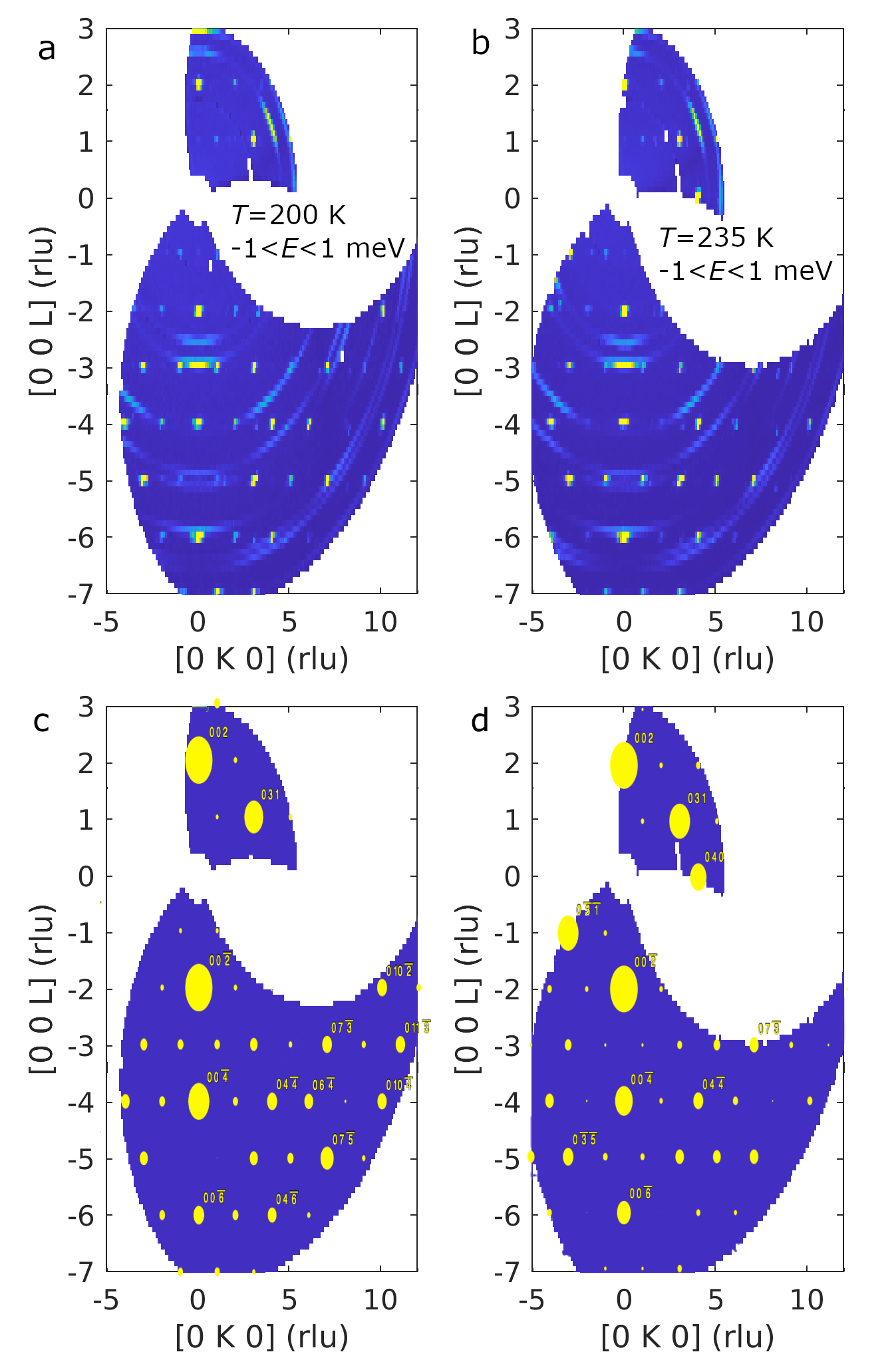}
\caption{Experimental $(0KL)$ reciprocal space maps for ammonium sulfate at the elastic line $-1<E<1$~meV at (a) $T=200$~K, and (b) $T=235$~K showing the sample Bragg peaks, plus streaks associated with elastic scattering from the aluminium sample mount and tails of the CCR. These can be compared to simulated single crystal diffraction maps for the (c)~$Pna2_1$ and (d) $Pnam$ phases; here the spot size represents the peak intensity.}\label{figqqslice}
\end{figure}

Using Horace\cite{Horace} we combined our individual data runs to construct the four dimensional $S(\mathbf{Q},\omega)$ dataset at $T=5$ and $200$~K in the $Pna2_1$ phase, and at $T=235$~K in the high temperature $Pnam$ phase. By looking at the elastic line (Fig.~\ref{figqqslice}) we are able to establish that there are no changes in the systematic absences between the two phases, just modifications of the relative peak intensities. The Bragg peaks are well localised indicating that the mosaic spread of the individual crystals is of reasonable quality. The streaks seen in the data arise from background scattering from the aluminium in the sample mount and from the tails of the CCR.

\begin{figure}[h]
    \centering
    \includegraphics[width=0.9\linewidth]{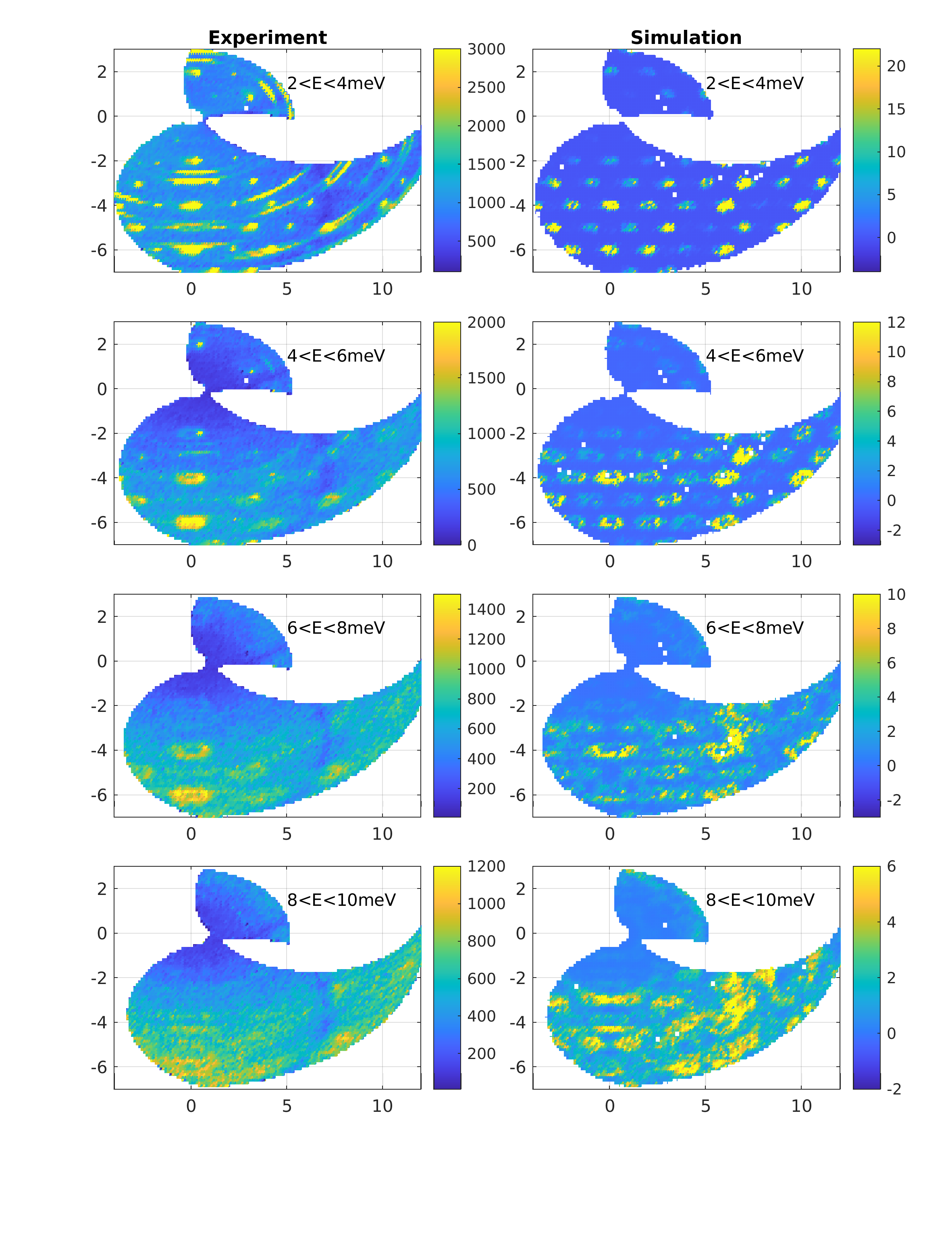}
    \caption{Data and simulations for inelastic energy $(0KL)$ slices of $Pna2_1$ \amsulfd\ through the acoustic modes at $T=200$~K.}\label{fig0KL}
  \end{figure}

A clear comparison between data and simulation in the low temperature phase can be made by looking at narrow inelastic energy slices in the $(0KL)$ plane (Fig.~\ref{fig0KL}). These Euphonic simulations include the instrumental resolution function which broadens the signal output from our CASTEP calculations. It is clear that our simulations overestimate the group velocity of the acoustic modes, such that the rings of scattering open up further at lower energy transfers than in the data slices.

\begin{figure}
  \centering
  \includegraphics[width=0.85\linewidth]{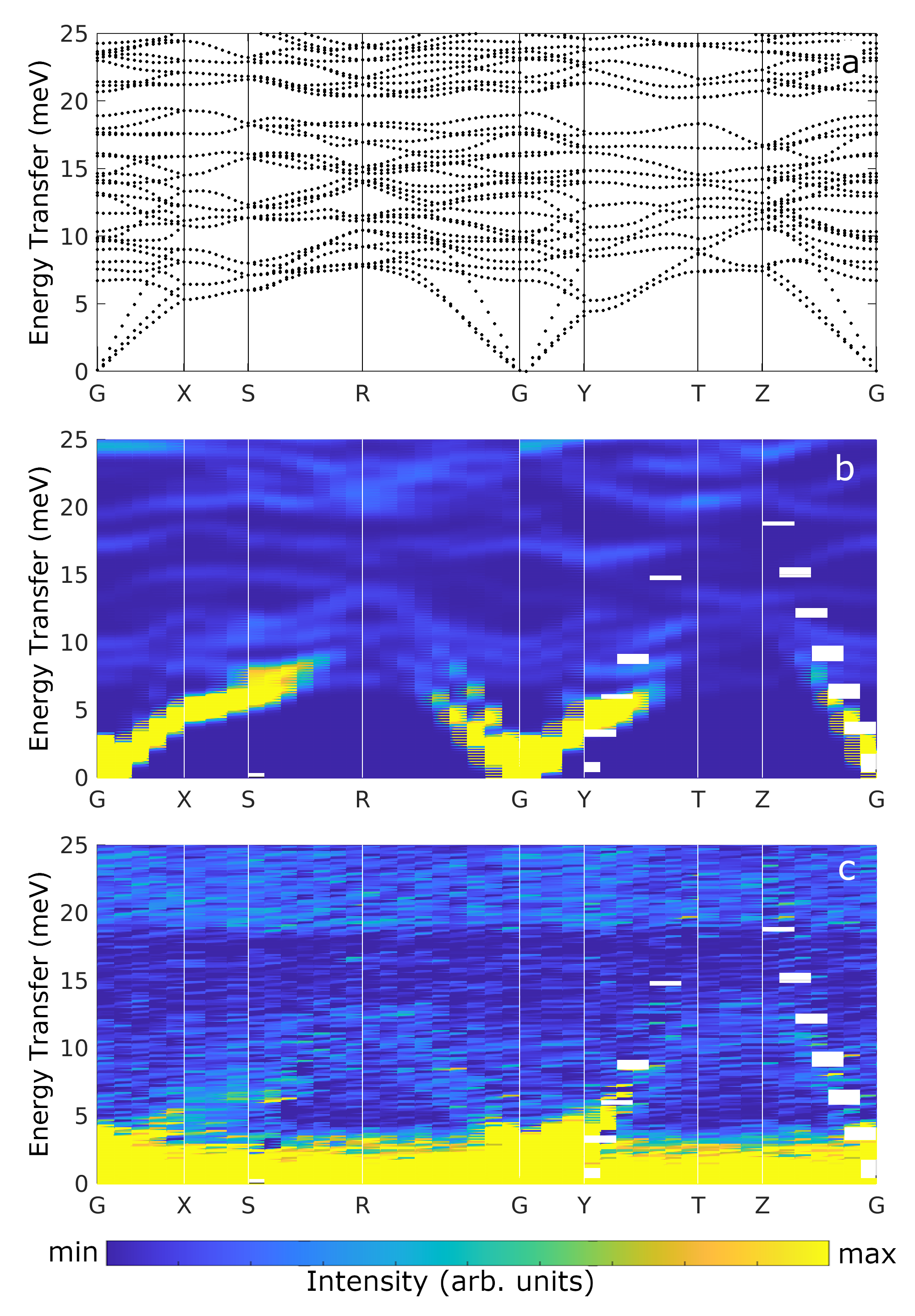}
  \caption{Phonon dispersion of \amsulfd\ in the low temperature $Pna2_1$ phase: (a) DFT calculations, (b) Euphonic calculations, and (c) data measured at $T=5$~K.}\label{figspagPna21}
\end{figure}

We further mapped out the phonon dispersion of single crystal \amsulfd\ in the low-temperature phase along the high symmetry directions in the Brillouin zone (Fig.~\ref{figspagPna21}). At base temperature, acoustic modes are observed emerging in the data from the Gamma points, along with a gap in the data at around 17 meV across the whole Brillouin Zone (Fig.~\ref{figspagPna21}(c)). The DFT calculations for $Pna2_1$ show far more modes (Fig.~\ref{figspagPna21}(a)) than are visible in our experimental data, but once the Euphonic code \cite{Euphonic} has been used to apply the neutron weighting (Fig.~\ref{figspagPna21}(b)) a good agreement with the data is seen for the low energy dispersive acoustic modes, with the gradient variation across the BZ reproduced particularly well. The higher energy modes are flatter in both the data and simulation, but the energy transfer corresponding to the gap in the scattering is overestimated in the simulation. 

\begin{figure}
  \centering
  \includegraphics[width=0.85\linewidth]{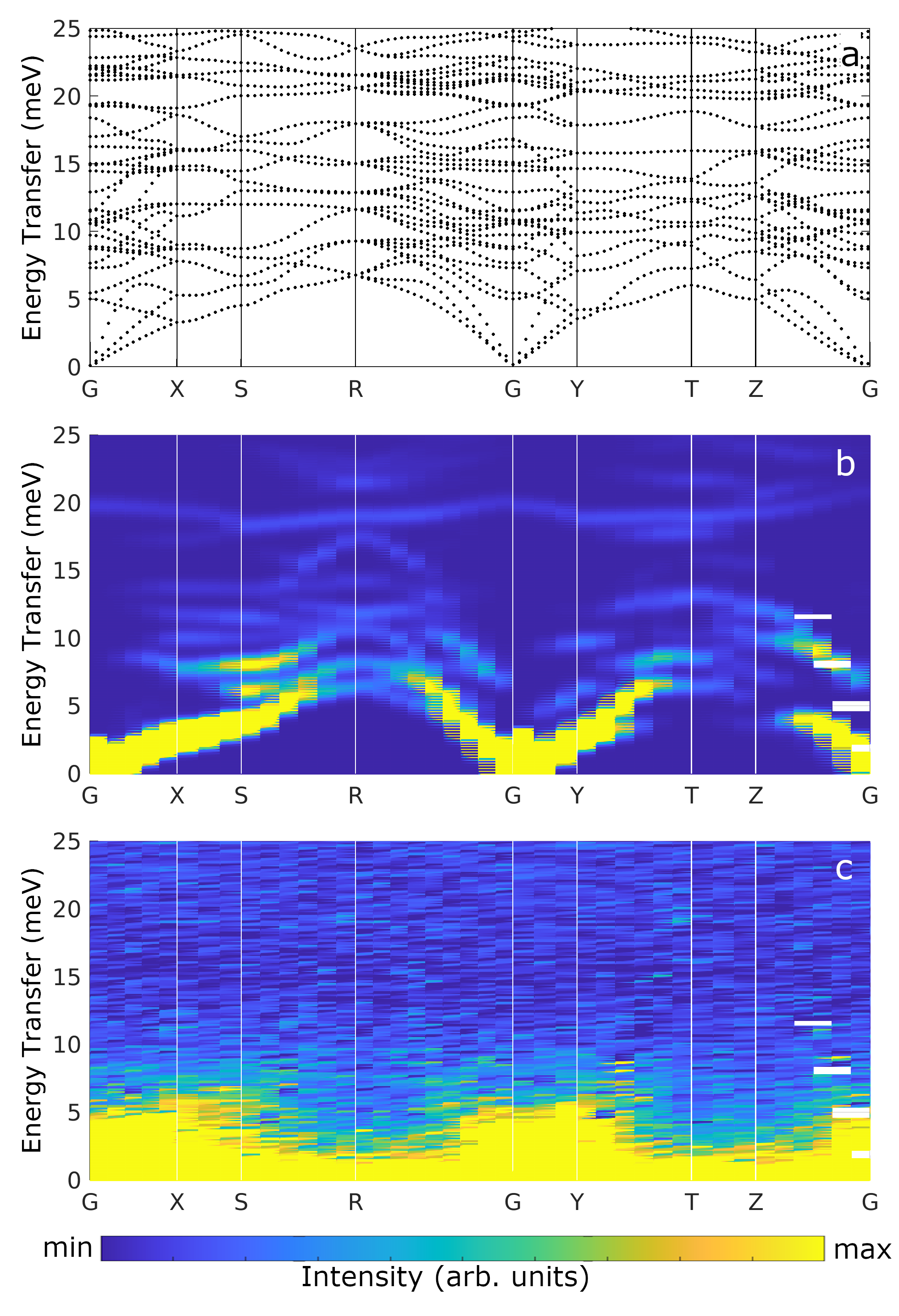}
  \caption{Phonon dispersion of \amsulfd\ in the high temperature $Pnam$ phase: (a) DFT calculations, (b) Euphonic calculations, and (c) data measured at $T=235$~K.}\label{figspagPnam}
\end{figure}

At $T=235$~K, the inelastic neutron scattering data are harder to interpret (Fig.~\ref{figspagPnam}) as the phonons have broadened considerably and the multiphonon scattering increases the background at high temperatures, resulting in few clear features, making a comparison with simulation difficult. Both the DFT calculations (Fig.~\ref{figspagPnam}(b)) and the Euphonic simulation (Fig.~\ref{figspagPnam}(c)) show that the lowest energy acoustic mode has softened, and there are modes crossing the energy gap that is present in $Pna2_1$, and the experimental data are consistent with this.

%

\section{Polycrystalline \amsulfd}
\subsection{Preliminary results}
\begin{figure}[h]
  \centering
  \includegraphics[width=0.55\textwidth]{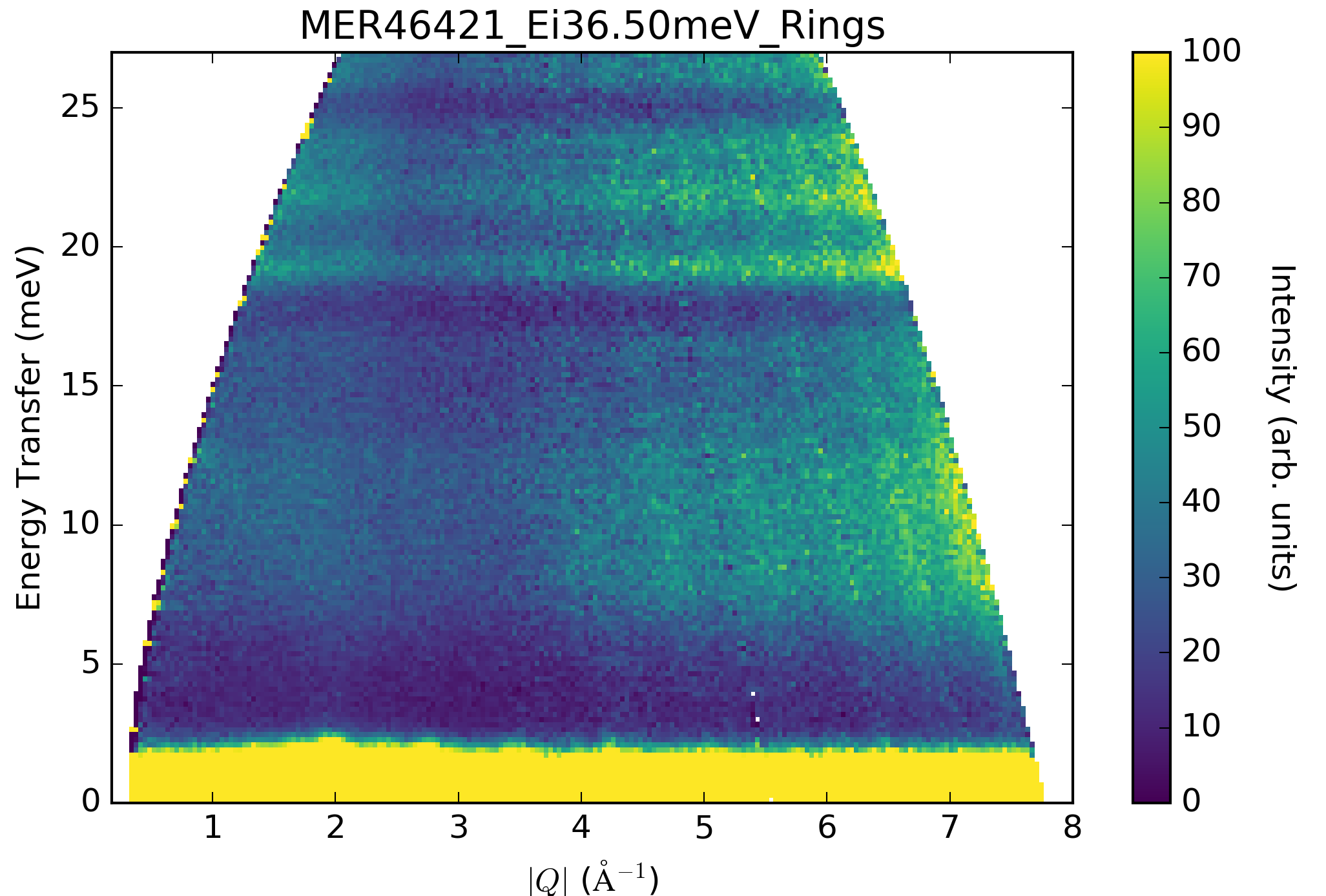}
  \caption{Colormap for the inelastic scattering from \amsulfd\ powder at $T=10$~K and ambient pressure as a function of momentum transfer and energy transfer.}\label{fig2Dslice}
\end{figure}

Figure~\ref{fig2Dslice} shows a typical inelastic neutron scattering data set for polycrystalline \amsulfd\ measured at base temperature and ambient pressure as a function of momentum $|\mathbf{Q}|$ and energy transfer for an incident energy of $36.5$~meV in the TAV6 clamp cell. The colorscale represents the measured signal intensity. Up to $\sim2$~meV, the strong signal seen is elastic scattering from \amsulfd\ and the clamp cell, due to the truncation of the intensity scale. Below $17$~meV the inelastic data is quite diffuse, due to powder averaging over dispersive acoustic modes. There is a gap in the scattering intensity at $18$~meV, and then a series of flat bands of scattering up to $24$~meV, corresponding to much less dispersive ammonium ``rattling'' modes. The diagonal striations modulating the data arise from an imperfect oscillation of the mini radial collimator fitted over the pressure cell.

\subsection{DFT phonon mode assignment}

Our CASTEP\cite{Castep_Clark,Castep_Refson} calculations allow us to assign the different phonon modes in the two different phases. Tables~\ref{TabPna21} and \ref{TabPnam} list the first eighty modes in $Pna2_1$ and $Pnam$ respectively, giving the energy at the Gamma point and the mode classification.

\newpage
\pagestyle{empty}

\begin{table}
  \centering
  \begin{minipage}[t]{0.325\textwidth}\scriptsize
    \vspace{0pt}
    \begin{tabular}{|m{6mm}|m{12mm}|m{18mm}|}
        \hline
        Mode & E (meV) & Motion Type\\
        \hline
        1 & 0.0 &\multirow{3}{18mm}{Zero energy translation}\\
        2 & 0.0 & \\
        3 & 0.0 & \\
        \hline
        4 & 6.4 & \multirow{45}{18mm}{Whole body motion: ND$_4$ and SO$_4$ translations coupled with SO$_4$ rotations} \\
        5 & 7.1 & \\
        6 & 7.3 & \\
        7 & 8.5 & \\
        8 & 8.9 & \\
        9 & 9.7 & \\
        10 & 9.8 & \\
        11 & 10.0 & \\
        12 & 11.2 & \\
        13 & 12.2 & \\
        14 & 12.5 & \\
        15 & 13.2 & \\
        16 & 13.2 & \\
        17 & 13.6 & \\
        18 & 14.0 & \\
        19 & 15.2 & \\
        20 & 15.2 & \\
        21 & 16.7 & \\
        22 & 16.8 & \\
        23 & 17.2 & \\
        24 & 18.6 & \\
        25 & 19.4 & \\
        26 & 20.1 & \\
        27 & 20.6 & \\
        28 & 21.6 & \\
        29 & 22.1 & \\
        30 & 22.2 & \\
        31 & 22.4 & \\
        32 & 22.6 & \\
        33 & 22.9 & \\
        34 & 24.4 & \\
        35 & 24.9 & \\
        36 & 25.7 & \\
        37 & 26.0 & \\
        38 & 26.3 & \\
        39 & 26.4 & \\
        40 & 27.0 & \\
        41 & 27.0 & \\
        42 & 27.9 & \\
        43 & 28.7 & \\
        44 & 28.9 & \\
        45 & 29.2 & \\
        46 & 29.7 & \\
        47 & 29.7 & \\
        48 & 33.1 & \\
        \hline
        49 & 34.3 & \multirow{12}{18mm}{ND$_4$ rotations}\\
        50 & 34.3 & \\
        51 & 34.6 & \\
        52 & 34.7 & \\
        53 & 39.8 & \\
        54 & 39.9 & \\
        55 & 40.1 & \\
        56 & 40.2 & \\
        57 & 41.8 & \\
        58 & 42.3 & \\
        59 & 42.9 & \\
        60 & 43.5 & \\

   \hline
    \end{tabular}
  \end{minipage}
  \begin{minipage}[t]{0.325\textwidth}\scriptsize
    \vspace{0pt}
      \begin{tabular}{|m{6mm}|m{12mm}|m{18mm}|}
        \hline
        Mode & E (meV) & Motion Type\\
        \hline
        61 & 43.8 & \multirow{12}{18mm}{ND$_4$ rotations}\\
        62 & 44.1 & \\
        63 & 44.2 & \\
        64 & 44.3 & \\
        65 & 45.9 & \\
        66 & 46.0 & \\
        67 & 46.0 & \\
        68 & 46.1 & \\
        69 & 48.3 & \\
        70 & 48.8 & \\
        71 & 48.9 & \\
        72 & 49.0 & \\
        \hline
        73 & 52.5 & \multirow{8}{18mm}{SO$_4$ $\nu_2$}\\
        74 & 53.1 & \\
        75 & 53.2 & \\
        76 & 53.2 & \\
        77 & 55.2 & \\
        78 & 55.5 & \\
        79 & 55.6 & \\
        80 & 55.9 & \\
        \hline
        81 & 71.6 & \multirow{12}{18mm}{SO$_4$ $\nu_4$}\\
        82 & 71.8 & \\
        83 & 71.8 & \\
        84 & 72.2 & \\
        85 & 72.3 & \\
        86 & 72.7 & \\
        87 & 72.8 & \\
        88 & 72.9 & \\
        89 & 73.7 & \\
        90 & 74.0 & \\
        91 & 74.2 & \\
        92 & 74.7 & \\
        \hline
        93 & 112.6 & \multirow{4}{18mm}{SO$_4$ $\nu_1$}\\
        94 & 112.7 & \\
        95 & 112.9 & \\
        96 & 112.9 & \\
        \hline
        97 & 121.6 & \multirow{24}{18mm}{SO$_4$ $\nu_3$ coupled with ND$_4$ $\nu_4$}\\
        98 & 121.8 & \\
        99 & 122.9 & \\
        100 & 124.2 & \\
        101 & 125.6 & \\
        102 & 126.4 & \\
        103 & 128.2 & \\
        104 & 128.6 & \\
        105 & 129.2 & \\
        106 & 129.3 & \\
        107 & 129.7 & \\
        108 & 131.1 & \\
        109 & 131.7 & \\
        110 & 132.0 & \\
        111 & 132.3 & \\
        112 & 132.5 & \\
        113 & 132.8 & \\
        114 & 133.3 & \\
        115 & 133.6 & \\
        116 & 134.2 & \\
        117 & 135.2 & \\
        118 & 135.9 & \\
        119 & 136.0 & \\
        120 & 136.8 & \\
        \hline
    \end{tabular}
  \end{minipage}
  \begin{minipage}[t]{0.325\textwidth}\scriptsize
    \vspace{0pt}
    \begin{tabular}{|m{6mm}|m{12mm}|m{18mm}|}
        \hline
        Mode & E (meV) & Motion Type\\
        \hline
        121 & 136.9 & \multirow{12}{18mm}{SO$_4$ $\nu_3$ coupled with ND$_4$ $\nu_4$}\\
        122 & 137.1 & \\
        123 & 137.4 & \\
        124 & 138.7 & \\
        125 & 138.7 & \\
        126 & 139.0 & \\
        127 & 139.7 & \\
        128 & 139.8 & \\
        129 & 139.9 & \\
        130 & 141.4 & \\
        131 & 141.4 & \\
        132 & 141.6 & \\
        \hline
        133 & 148.2 & \multirow{16}{18mm}{ND$_4$ $\nu_2$}\\
        134 & 148.2 & \\
        135 & 148.2 & \\
        136 & 148.5 & \\
        137 & 149.1 & \\
        138 & 149.2 & \\
        139 & 149.2 & \\
        140 & 149.6 & \\
        141 & 149.8 & \\
        142 & 150.5 & \\
        143 & 150.8 & \\
        144 & 151.2 & \\
        145 & 151.9 & \\
        146 & 152.0 & \\
        147 & 152.1 & \\
        148 & 152.4 & \\
        \hline
        149 & 266.7 & \multirow{8}{18mm}{ND$_4$ $\nu_1$}\\
        150 & 266.9 & \\
        151 & 267.2 & \\
        152 & 267.8 & \\
        153 & 268.4 & \\
        154 & 268.5 & \\
        155 & 268.5 & \\
        156 & 269.1 & \\
        \hline
        157 & 274.4 & \multirow{24}{18mm}{ND$_4$ $\nu_3$}\\
        158 & 274.5 & \\
        159 & 274.7 & \\
        160 & 274.9 & \\
        161 & 276.8 & \\
        162 & 277.9 & \\
        163 & 278.1 & \\
        164 & 278.9 & \\
        165 & 281.1 & \\
        166 & 281.1 & \\
        167 & 281.4 & \\
        168 & 285.5 & \\
        169 & 286.8 & \\
        170 & 287.3 & \\
        171 & 287.5 & \\
        172 & 288.9 & \\
        173 & 291.2 & \\
        174 & 291.9 & \\
        175 & 292.7 & \\
        176 & 293.6 & \\
        177 & 307.9 & \\
        178 & 307.9 & \\
        179 & 308.3 & \\
        180 & 309.4 & \\
        \hline
      \end{tabular}
  \end{minipage}
  \caption{Table of the phonon modes at the $\Gamma$-point in \amsulfd\ in the low temperature $Pna2_1$ phase.}\label{TabPna21}
\end{table}

\begin{table}
  \centering
  \begin{minipage}[t]{0.325\textwidth}\scriptsize
    \vspace{0pt}
    \begin{tabular}{|m{6mm}|m{12mm}|m{18mm}|}
        \hline
        Mode & E (meV) & Motion Type\\
        \hline
        1 & -17.7 &\multirow{4}{18mm}{ND$_4$ libration: N2 about b}\\
        2 & -16.7 & \\
        3 & -16.1 & \\
        4 & -15.9 & \\
        \hline
        5 & -7.6 & \multirow{3}{18mm}{ND$_4$ libration: N1 about a} \\
        6 & -6.8 & \\
        7 & -2.9 & \\
        \hline
        8 & 0.0 & \multirow{3}{18mm}{Zero energy translation} \\
        9 & 0.0 & \\
        10 & 0.0 & \\
        \hline
        \multirow{1}{6mm}{11} & \multirow{1}{12mm}{8.4} & ND$_4$ lib: N1a\\
        \hline
        12 & 6.0 & \multirow{49}{18mm}{Whole body motion: ND$_4$ and SO$_4$ coupled translations and rotations}\\
        13 & 6.9 & \\
        14 & 7.6 & \\
        15 & 7.7 & \\
        16 & 7.8 & \\
        17 & 8.0 & \\
        18 & 9.3 & \\
        19 & 10.0 & \\
        20 & 11.3 & \\
        21 & 12.5 & \\
        22 & 15.1 & \\
        23 & 15.5 & \\
        24 & 15.6 & \\
        25 & 17.5 & \\
        26 & 17.8 & \\
        27 & 17.9 &\\
        28 & 18.1 & \\
        29 & 19.5 & \\
        30 & 19.6 & \\
        31 & 19.7 & \\
        32 & 21.0 & \\
        33 & 21.2 & \\
        34 & 21.6 & \\
        35 & 22.1 & \\
        36 & 22.2 & \\
        37 & 22.3 & \\
        38 & 22.4 & \\
        39 & 22.7 & \\
        40 & 22.8 & \\
        41 & 22.8 & \\
        42 & 24.2 & \\
        43 & 24.8 & \\
        44 & 25.5 & \\
        45 & 25.5 & \\
        46 & 25.6 & \\
        47 & 25.9 & \\
        48 & 26.0 & \\
        49 & 26.1 &\\
        50 & 26.8 & \\
        51 & 27.2 & \\
        52 & 29.7 & \\
        53 & 29.9 & \\
        54 & 30.9 & \\
        55 & 31.0 & \\
        56 & 33.8 & \\
        57 & 34.2 & \\
        58 & 34.5 & \\
        59 & 34.6 & \\
        60 & 34.6 & \\
      \hline
    \end{tabular}
  \end{minipage}
  \begin{minipage}[t]{0.32\textwidth}
    \vspace{0pt}\scriptsize
      \begin{tabular}{|m{6mm}|m{12mm}|m{18mm}|}
        \hline
        Mode & E (meV) & Motion Type\\
        \hline
        61 & 34.7 & \multirow{8}{18mm}{Whole body motion: ND$_4$ and SO$_4$ coupled translations and rotations}\\
        62 & 34.8 & \\
        63 & 35.3 & \\
        64 & 35.6 & \\
        65 & 35.7 & \\
        66 & 35.9 & \\
        67 & 36.1 & \\
        68 & 37.7 & \\
        \hline
        69 & 46.0 & \multirow{4}{18mm}{ND$_4$ libration: N1 about c}\\
        70 & 46.4 & \\
        71 & 47.4 & \\
        72 & 47.6 & \\
        \hline
        73 & 52.7 & \multirow{8}{18mm}{SO$_4$ $\nu_2$}\\
        74 & 52.9 & \\
        75 & 52.9 & \\
        76 & 52.9 & \\
        77 & 53.2 & \\
        78 & 53.5 & \\
        79 & 53.5 & \\
        80 & 54.6 & \\
        \hline
        81 & 71.0 & \multirow{12}{18mm}{SO$_4$ $\nu_4$}\\
        82 & 71.1 & \\
        83 & 71.2 & \\
        84 & 71.6 & \\
        85 & 71.6 & \\
        86 & 71.7 & \\
        87 & 71.7 & \\
        88 & 72.0 & \\
        89 & 74.1 & \\
        90 & 74.7 & \\
        91 & 74.9 & \\
        92 & 75.3 & \\
        \hline
        93 & 113.9 & \multirow{4}{18mm}{SO$_4$ $\nu_1$}\\
        94 & 114.0 & \\
        95 & 114.0 & \\
        96 & 114.0 & \\
        \hline
        97 & 122.4 & \multirow{24}{18mm}{SO$_4$ $\nu_3$ coupled with ND$_4$ $\nu_4$}\\
        98 & 122.9 &\\
        99 & 123.3 &\\
        100 & 124.1 &\\
        101 & 124.6 &\\
        102 & 124.6 &\\
        103 & 124.9 &\\
        104 & 125.2 &\\
        105 & 125.5 &\\
        106 & 128.1 &\\
        107 & 129.4 &\\
        108 & 129.5 &\\
        109 & 130.1 &\\
        110 & 130.3 &\\
        111 & 130.6 &\\
        112 & 130.7 &\\
        113 & 131.3 &\\
        114 & 131.4 &\\
        115 & 133.1 &\\
        116 & 133.7 &\\
        117 & 133.8 &\\
        118 & 133.9 &\\
        119 & 133.9 &\\
        120 & 134.7 &\\
      \hline
    \end{tabular}
  \end{minipage}
  \begin{minipage}[t]{0.32\textwidth}
    \vspace{0pt}\scriptsize
    \begin{tabular}{|m{6mm}|m{12mm}|m{18mm}|}
        \hline
        Mode & E (meV) & Motion Type\\
        \hline
        121 & 134.8 & \multirow{12}{18mm}{SO$_4$ $\nu_3$ coupled with ND$_4$ $\nu_4$}\\
        122 & 135.0 &\\
        123 & 135.1 &\\
        124 & 137.1 &\\
        125 & 137.4 &\\
        126 & 138.4 &\\
        127 & 140.4 &\\
        128 & 140.9 &\\
        129 & 141.0 &\\
        130 & 141.3 &\\
        131 & 141.9 &\\
        132 & 142.4 &\\
        \hline
        133 & 145.6 & \multirow{16}{18mm}{ND$_4$ $\nu_2$}\\
        134 & 146.2 &\\
        135 & 146.4 &\\
        136 & 146.6 &\\
        137 & 147.2 &\\
        138 & 147.8 &\\
        139 & 148.5 &\\
        140 & 148.5 &\\
        141 & 148.7 &\\
        142 & 149.3 &\\
        143 & 149.5 &\\
        144 & 149.9 &\\
        145 & 150.5 &\\
        146 & 150.6 &\\
        147 & 150.7 &\\
        148 & 150.9 &\\
        \hline
        149 & 266.7 & \multirow{8}{18mm}{ND$_4$ $\nu_1$}\\
        150 & 267.5 &\\
        151 & 268.4 &\\
        152 & 269.6 &\\
        153 & 276.8 &\\
        154 & 276.8 &\\
        155 & 277.2 &\\
        156 & 277.4 &\\
        \hline
        157 & 288.0 & \multirow{24}{18mm}{ND$_4$ $\nu_3$}\\
        158 & 288.0 &\\
        159 & 288.5 &\\
        160 & 288.7 &\\
        161 & 290.6 &\\
        162 & 290.7 &\\
        163 & 292.7 &\\
        164 & 293.7 &\\
        165 & 293.7 &\\
        166 & 293.7 &\\
        167 & 293.7 &\\
        168 & 293.8 &\\
        169 & 295.3 &\\
        170 & 295.5 &\\
        171 & 295.5 &\\
        172 & 295.8 &\\
        173 & 295.9 &\\
        174 & 296.4 &\\
        175 & 298.1 &\\
        176 & 298.1 &\\
        177 & 313.1 &\\
        178 & 313.2 &\\
        179 & 314.4 &\\
        180 & 314.4 &\\
        \hline
      \end{tabular}
  \end{minipage}
  \caption{Table of the phonon modes at the $\Gamma$-point in \amsulfd\ in the high temperature $Pnam$ phase.}\label{TabPnam}
\end{table}

\clearpage
\pagestyle{plain}

\begin{figure}[h]
  \centering
  \includegraphics[width=0.45\linewidth]{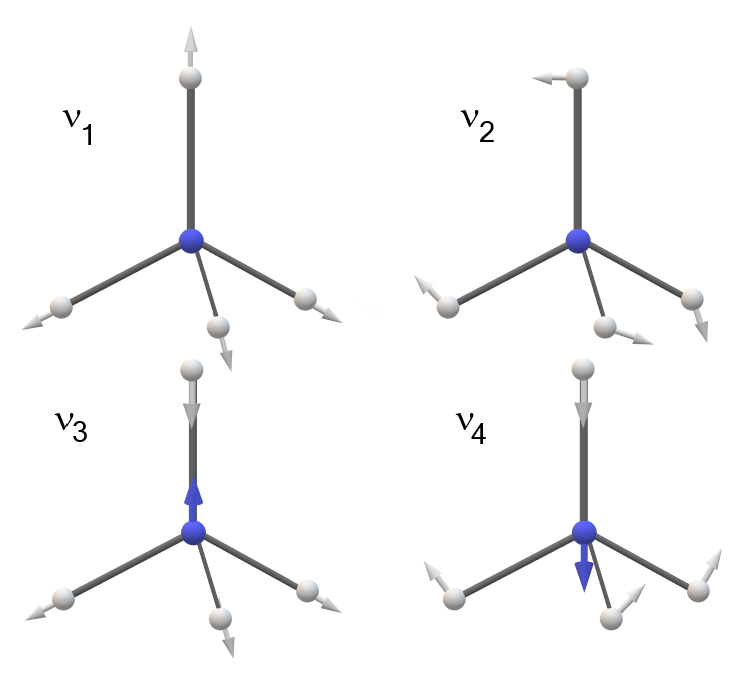}
  \caption{The normal modes of a free tetrahedral molecule.}\label{figtetra}
\end{figure}

The normal modes of the tetrahedral space group ($\nu_1-\nu_4$) referred to in the tables are shown in Fig.~\ref{figtetra}. The $\nu_1$ mode is the $A_1$ breathing mode. The $\nu_2$ mode is the doubly degenerate $E$ bending mode. The $\nu_3$ mode is the triply degenerate $T_2$ stretching mode. The $\nu_4$ mode is the triply degenerate $T_2$ bending mode.

\subsection{Temperature variation}
\begin{figure}
  \centering
  \includegraphics[width=0.5\linewidth]{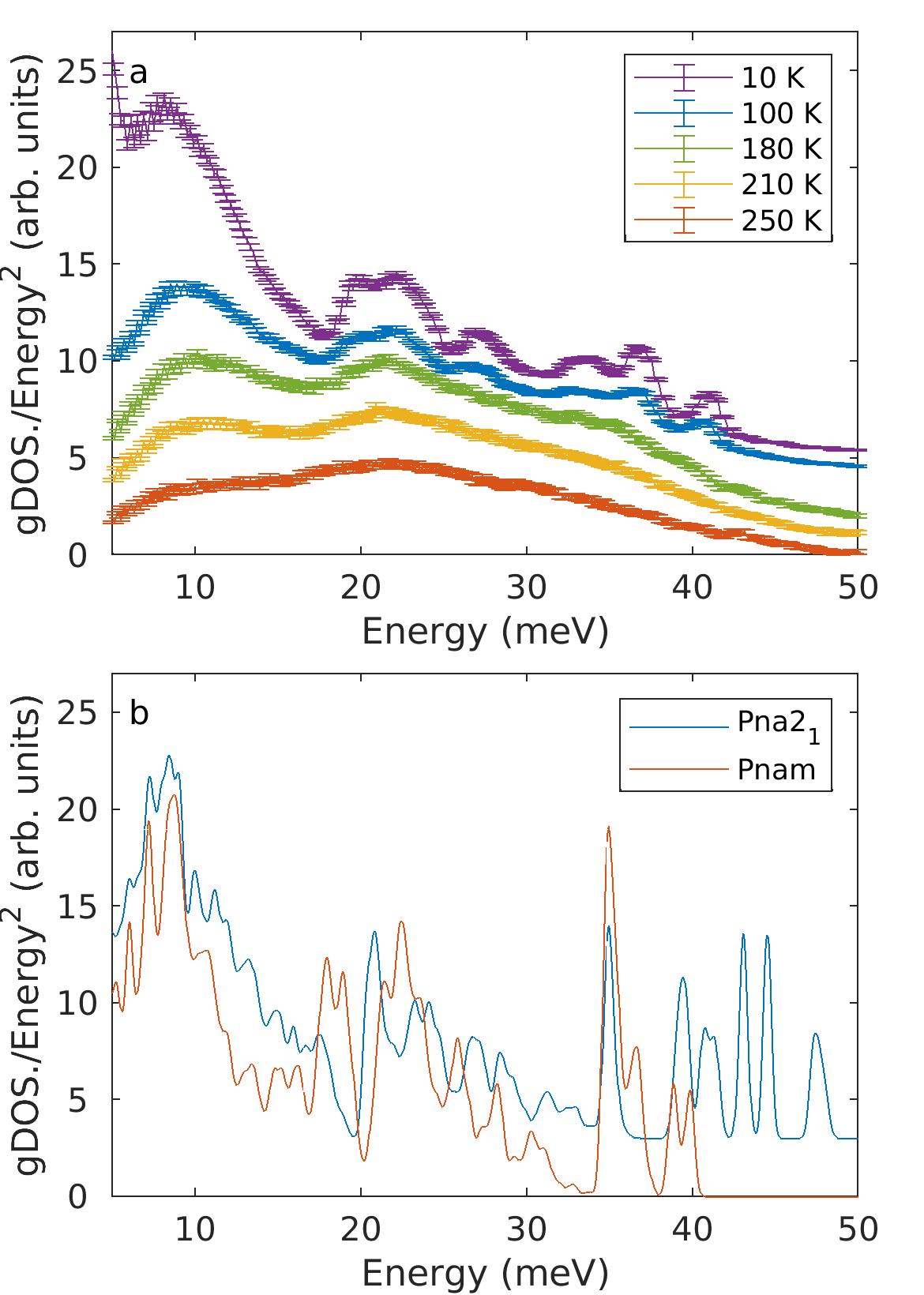}
  \caption{(a) The temperature dependence of the neutron weighted generalised phonon density of states of \amsulfd\ measured with $E_i=67$~meV divided by the energy transfer squared, and (b) a comparison of the simulated phonon density of states in the $Pna2_1$ and $Pnam$ phases divided by the energy transfer squared. The spectra have been offset vertically, and the data in (a) for $T=180-250$~K have been multiplied by two for clarity.}\label{figTdep}
\end{figure}

We tracked the evolution of the experimental inelastic scattering as a function of both temperature and pressure. Figure.~\ref{figTdep}(a) shows how the phonon density of states divided by energy transfer squared evolves as a function of temperature at ambient pressure; these data have been scaled by the Bose factor, demonstrating the broadening of the sharp features seen at base temperature. Regrettably, this reveals that there is no clear signature of a change in the density of states associated with the phase transition at $T=224$~K. It could perhaps be argued that the dip present at $E=18$~meV at base temperature is present all the way up to $T=210$~K, but then closes on entering the high temperature phase (Fig.~\ref{figTdep}(a)).
This would be consistent with simulations for $Pna2_1$ and $Pnam$ (Fig.~\ref{figTdep}(b)) performed at zero temperature, where, as noted in the previous section, the most obvious difference between the two is the lowering in energy of the ammonium librational modes. This results in a peak in the calculated gDOS at $18$~meV in $Pnam$, where there is a dip in $Pna2_1$.

\begin{figure}
  \centering
  \includegraphics[width=0.5\textwidth]{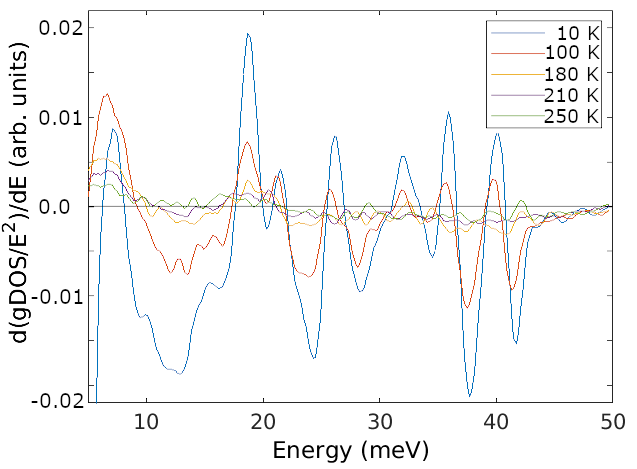}
  \caption{Numerical derivative of the generalised neutron-weighted density of states divided by energy transfer squared for \amsulfd\ measured at temperatures between 10 and 250~K.}\label{figdiff}
\end{figure}


In this figure, dividing by $Q^2$ was done to highlight the change at the phase transition from the presence of a minimum at 18~meV below to something more like an inversion point above. To emphasize this point, Fig.~\ref{figdiff} shows the numerical derivative of the data presented in Fig.~\ref{figTdep}, which was first smoothed using a gaussian filter. In the low temperature phase data there is a distinct maximum in the derivative at $17-18$~meV for each temperature, whereas no such clear maximum is present in the derivative at $T=250$~K in the high temperature phase.

\clearpage
\section{Theory of nuclear inelastic neutron scattering}
\subsection{Scattering function}
Since the neutron matter interaction is weak, the interaction matrix for nuclear scattering can be evaluated in the Born approximation to give the dynamical scattering function:
\begin{equation}\label{eqS}
  S(\mathbf{Q},\omega)=\frac{1}{2\pi\hbar}\sum_{i,j}\overline{b_ib_j}\int_{-\infty}^{\infty}\langle\exp(i\mathbf{Q}\cdot\left[\mathbf{r}_i(t)-\mathbf{r}_j(0)\right])\rangle\exp(-i\omega t) dt,
\end{equation}
where $\mathbf{Q}$ and $\hbar\omega$ are the change in momentum and energy of the neutron on scattering, $b$ is the nuclear scattering length, $t$ is the time and $\mathbf{r}$ are the coordinates of the scattering centres. The scattering can be separated into coherent and incoherent scattering, where coherent scattering provides information about the relative positions of the atoms and their collective dynamics, and incoherent scattering provides information about single particle dynamics. Assuming there is no correlation between the $b$ values of different nuclei:
\begin{align}
  \nonumber 
  &&&&\overline{b_ib_j}&= (\overline{b})^2, & i &\neq j&&& \\
  &&&&\overline{b_ib_j}&= \overline{b^2}, & i &= j,&&&
\end{align}
one can separate the scattering function into its coherent and incoherent parts:
\begin{align}
  S(\mathbf{Q},\omega)_{coh}&=\frac{(\overline{b})^2}{2\pi\hbar}\sum_{i,j}\int_{-\infty}^{\infty}\langle\exp(i\mathbf{Q}\cdot\left[\mathbf{r}_i(t)-\mathbf{r}_j(0)\right])\rangle\exp(-i\omega t) dt, \label{eqcoh}\\
  S(\mathbf{Q},\omega)_{inc}&=\frac{(\overline{b^2}-(\overline{b})^2)}{2\pi\hbar}\sum_{i}\int_{-\infty}^{\infty}\langle\exp(i\mathbf{Q}\cdot\left[\mathbf{r}_i(t)-\mathbf{r}_i(0)\right])\rangle\exp(-i\omega t) dt, \label{eqinc}
\end{align}
where the coherent cross section $\sigma_{coh}=4\pi(\overline{b})^2$ and the incoherent cross section is $\sigma_{inc}=4\pi(\overline{b^2}-(\overline{b})^2)$.

\subsection{Scattering by crystals}
In an ordered crystal, the instantaneous position of an atom $d$ is
\begin{equation}
  \mathbf{r}_{ld}(t)=\mathbf{l}+\mathbf{d}+\mathbf{u}_{ld}(t),
\end{equation}
where $\mathbf{l}$ is the vector from the origin to lattice point $l$, $\mathbf{d}$ connects the lattice point to the equilibrium position of the atom and $\mathbf{u}_{ld}(t)$ is the displacement from the average position. The summation over all nuclei can be separated into a sum over all lattice pairs $l$, $l'$ and a sum over all pairs of atoms in the unit cell $d$ and $d'$; and as we can treat the lattice as infinite, the sum over $l$ is the same for each $l'$, so we need only invlude one value of $l'$, say $l'=0$, and then multiply through by the total number of lattice points $N$. Thus:
\begin{equation}
  S(\mathbf{Q},\omega)=\frac{N}{2\pi\hbar}\sum_l\exp(i\mathbf{Q}\cdot\mathbf{l})\sum_{d,d'}\overline{b_db_{d'}}\exp(i\mathbf{Q}\cdot(\mathbf{d}-\mathbf{d}'))
   \times\int_{-\infty}^{\infty}\langle\exp\left(i\mathbf{Q}\cdot(\mathbf{u}_{ld}(t)-\mathbf{u}_{0d'})\right)\rangle\exp(-i\omega t) dt.
\end{equation}
The term in angled brackets can be rewritten as
\begin{equation}
  \exp\left(i\mathbf{Q}\cdot(\mathbf{u}_{ld}(t)-\mathbf{u}_{0d'})\right)=\exp(-\langle(\mathbf{Q}\cdot\mathbf{u}_d)^2/2\rangle)\times\exp(-\langle(\mathbf{Q}\cdot\mathbf{u}_{d'})^2/2\rangle)\times\exp(\langle[\mathbf{Q}\cdot\mathbf{u}_{ld}(t)][\mathbf{Q}\cdot\mathbf{u}_{0d'}]\rangle),
\end{equation}
where the first two product terms are the Debye-Waller factors, which have no dependence on $l$ since translational symmetry means that the mean squared displacement of an atom at a particular crystallographic site is the same in all unit cells. The third term can then be expanded in a power expansion:
\begin{equation}
\exp(\langle[\mathbf{Q}\cdot\mathbf{u}_{ld}(t)][\mathbf{Q}\cdot\mathbf{u}_{0d'}]\rangle)=\sum_{m=0}^\infty\frac{1}{m!}\langle[\mathbf{Q}\cdot\mathbf{u}_{ld}(t)][\mathbf{Q}\cdot\mathbf{u}_{0d'}]\rangle^m
\end{equation}
The $m$-th term relates to the cross-section for all $m$-phonon processes, such that the first term involves no phonons and is Bragg scattering, while $m=1$ gives the cross-section for all one-phonon processes. Thus the one phonon scattering function is:
\begin{align}\label{eqS1ph}
   \nonumber
   S(\mathbf{Q},\omega)&=\frac{N}{2\pi\hbar}\sum_l\exp(i\mathbf{Q}\cdot\mathbf{l})\sum_{d,d'}\overline{b_db_{d'}}\exp(i\mathbf{Q}\cdot(\mathbf{d}-\mathbf{d}'))\exp(-W_d(\mathbf{Q})\exp(-W_{d'}(\mathbf{Q}))\\
   &\times\int_{-\infty}^\infty\langle[\mathbf{Q}\cdot\mathbf{u}_{ld}(t)][\mathbf{Q}\cdot\mathbf{u}_{0d'}]\rangle\exp(-i\omega t)dt.
\end{align}

\subsection{The 1-phonon scattering function}
Assuming the interatomic forces are harmonic, the displacement $\mathbf{u}_{ld}(t)$ can be expressed as the sum of a set of normal modes to give
\begin{equation}
\mathbf{Q}\cdot\mathbf{u}_{ld}(t)=\left(\frac{\hbar}{2m_dN}\right)^{1/2}\sum_s\frac{\mathbf{Q}\cdot\mathbf{e}_{ds}}{\sqrt{\omega_s}}\left[a_s\exp(i(\mathbf{q}\cdot\mathbf{l}-\omega_st))+a^\dag_{-s}\exp(-i(\mathbf{q}\cdot\mathbf{l}-\omega_{-s}t))\right],
\end{equation}
where $\mathbf{q}$ is the wavevector of the mode, $\omega_s$ is the frequency of mode $s$, and $\mathbf{e}_s$ is its polarization vector. The sum over $s$ is over the $N$ values of $q$ in the first Brillouin zone, and $m_d$ is the mass of the atom. $\hat{a}_s$ and $\hat{a}^\dag_s$ are the annihilation and creation operators for mode $s$ with explicit time dependence:
\begin{align}
\hat{a}(t) &=a\exp(-i\omega t),\\
\hat{a}^\dag(t) &=a^\dag\exp(i\omega t),
\end{align}
and
\begin{align}
\langle a^\dag a\rangle&=n(\omega),\\
\langle aa^\dag\rangle&=n(\omega+1),
\end{align}
where $n(\omega)$ is the average occupation number of boson quasiparticles in a state with energy $\hbar\omega$.
The scattering function in Eq.~\eqref{eqS1ph} can be rewritten as
\begin{align}\label{eqS1phnorm}
   \nonumber
   S(\mathbf{Q},\omega)^{1\,\mathrm{ph}}&=\sum_l\exp(i\mathbf{Q}\cdot\mathbf{l})\sum_{d,d'}\frac{\overline{b_db_{d'}}}{4\pi\sqrt{m_dm_{d'}}}\exp(i\mathbf{Q}\cdot(\mathbf{d}-\mathbf{d'}))\exp(-W_d(\mathbf{Q})\exp(-W_{d'}(\mathbf{Q}))\sum_{s,s'}\frac{(\mathbf{Q}\cdot\mathbf{e}_{ds})(\mathbf{Q}\cdot\mathbf{e}_{d's'})}{\sqrt{\omega_s\omega_{s'}}}\\
   &\times\int_{-\infty}^\infty\left[\langle n_s\rangle\exp(-i(\omega+\omega_s)t)\exp(i\mathbf{q}\cdot\mathbf{l}) +
   \langle n_{-s'}+1\rangle\exp(-i(\omega-\omega_{-s'})t)\exp(-i\mathbf{q}\cdot\mathbf{l})\right]  dt.
\end{align}
Finally, noting that $\omega_s=\omega_{-s}$ and that normal modes are orthogonal, such that only terms with $s=s'$ survive, applying the time integral and the lattice summation give
\begin{align}\label{eqS1phdelta}
  \nonumber
  S(\mathbf{Q},\omega)^{1\,\mathrm{ph}}&= \frac{(2\pi)^3}{v_0}\sum_\mathbf{G}\sum_s\sum_{d,d'}\frac{\overline{b_db_{d'}}}{\sqrt{m_dm_{d'}}}\exp(i\mathbf{Q}\cdot(\mathbf{d}-\mathbf{d}'))\exp(-W_d(\mathbf{Q}))\exp(-W_{d'}(\mathbf{Q})) \\
  &\times\frac{(\mathbf{Q}\cdot\mathbf{e}_{ds})(\mathbf{Q}\cdot\mathbf{e}_{d's})}{2\omega_s}\left[\langle n_s\rangle\delta(\mathbf{Q}+\mathbf{q}-\mathbf{G})\delta(\omega+\omega_s)+\langle n_s+1\rangle\delta(\mathbf{Q}-\mathbf{q}-\mathbf{G})\delta(\omega-\omega_s)\right],
\end{align}
where $\mathbf{G}$ is a reciprocal lattice vector and $v_0$ is the volume of the unit cell. The delta functions dictate that scattering only occurs when momentum and energy are conserved, i.e. when $\mathbf{Q}$ and $\omega$ satisfy the conditions:
\begin{equation}
\mathbf{Q}=\mathbf{G}\pm\mathbf{Q} \qquad\mathrm{and}\qquad \omega=\pm\omega_s,
\end{equation}
where $\mathbf{q}$ is defined in the first Brillouin zone. The first term in the square brackets of Eq.~\eqref{eqS1phdelta} corresponds to phonon annihilation leading to an increase in the kinetic energy of the neutron. The second term in the square brackets of Eq.~\eqref{eqS1phdelta} corresponds to the neutron creating a phonon. These processes are known as phonon absorption and emission respectively.

\subsection{The coherent 1-phonon scattering function}
Equation~\eqref{eqS1phdelta} can now be combined with Eq.~\eqref{eqcoh} to obtain the 1-phonon coherent scattering function:
\begin{align}\label{eqS1phcoh}
  \nonumber
  S(\mathbf{Q},\omega)^{1\,\mathrm{ph}}_{\mathrm{coh}}&= \frac{(2\pi)^3}{v_0}\sum_\mathbf{G}\sum_s|G_s(\mathbf{Q})|^2\\
  &\times\frac{1}{2\omega_s}\left[\langle n_s\rangle\delta(\mathbf{Q}+\mathbf{q}-\mathbf{G})\delta(\omega+\omega_s)+\langle n_s+1\rangle\delta(\mathbf{Q}-\mathbf{q}-\mathbf{G})\delta(\omega-\omega_s)\right],
\end{align}
where $G_s$ is the phonon structure factor for mode $s$:
\begin{equation}\label{eqphSF}
  G_s(\mathbf{Q})=\sum_d\frac{\overline{b}_d}{\sqrt{m_d}}\mathbf{Q}\cdot\mathbf{e}_{ds}\exp(-W_d)\exp(i\mathbf{Q}\cdot\mathbf{d}).
\end{equation}

\subsection{The incoherent 1-phonon scattering function}
Equivalently the 1-phonon incoherent scattering function can be obtained by combining Eq.~\eqref{eqS1phnorm} with Eq.~\eqref{eqinc}, simplifying considerably since $d=d'$:
\begin{align}\label{eqS1phinc}
  \nonumber
  S(\mathbf{Q},\omega)^{1\,\mathrm{ph}}_{\mathrm{inc}}&= \sum_s\sum_d\frac{(\sigma_{\mathrm{inc}})_d}{4\pi}\frac{1}{m_d}|\mathbf{Q}\cdot{e}_{ds}|^2\exp(-2W_d)\\
  &\times\frac{1}{2\omega_s}\left[\langle n_s\rangle\delta(\omega+\omega_s)+\langle n_s+1\rangle\delta(\omega-\omega_s)\right],
\end{align}
revealing that there is no $\delta$-function in $\mathbf{Q}$, or any interference term for different atoms in the unit cell, and so $S(\mathbf{Q},\omega)^{1\,\mathrm{ph}}_{\mathrm{inc}}$ varies continuously with $\mathbf{Q}$, and only exhibits a weak dependence on $\mathbf{Q}$ due to the polarisation and Debye-Waller factors.

As all modes of frequency $\omega$ will contribute to $S(\mathbf{Q},\omega)^{1\,\mathrm{ph}}_{\mathrm{inc}}$ at a given $\mathbf{Q}$ and $\omega$, we use the vibrational density of states $g(\omega)$ to replace the sum over the normal modes by an integral over $g(\omega)$ to give:
\begin{align}\label{eqS1phinc2}
  S(\mathbf{Q},\omega)^{1\,\mathrm{ph}}_{\mathrm{inc}}&= \sum_d\frac{(\sigma_{\mathrm{inc}})_d}{4\pi}\frac{1}{m_d}\exp(-2W_d)\int_{-\infty}^{\infty}g(\omega')\langle|\mathbf{Q}\cdot{e}_{ds}|^2\rangle_{\omega'} \\
  & \times\frac{1}{2\omega'}\left[\langle n(\omega')\rangle\delta(\omega+\omega')+\langle n(\omega')+1\rangle\delta(\omega-\omega')\right],
\end{align}
where $\langle|\mathbf{Q}\cdot{e}_{ds}|^2\rangle_{\omega'}$ is the value of $|\mathbf{Q}\cdot{e}_{ds}|^2$ averaged over all modes of frequency $\omega'$. Using the identity
\begin{equation}\label{eqident}
  n(-\omega)+1=-n(\omega),
\end{equation}
the integral can be evaluated to give
\begin{equation}\label{eqS1phinc3}
  S(\mathbf{Q},\omega)^{1\,\mathrm{ph}}_{\mathrm{inc}}= \sum_d\frac{(\sigma_{\mathrm{inc}})_d}{4\pi}\frac{1}{m_d}\exp(-2W_d)\langle|\mathbf{Q}\cdot{e}_{ds}|^2\rangle_{\omega}
  \frac{g(\omega)}{2\omega}\left[n(\omega)+1\right].
\end{equation}

\subsection{Incoherent Approximation}
Ideally, one would always prefer to obtain phonon dispersion relations from single crystal measurements, but this is time consuming and it is not always possible to synthesize single crystals, or their crystal habit makes them inconsistent with sample environment (e.g. in a pressure cell). Instead, useful information about the vibrational dynamics can still be determined through powder measurements of the density of states $g(\omega)$. In polycrystalline samples the single crystal grains are randomly oriented, such that, if anisotropy in the Debye-Waller can be neglected, the one-phonon incoherent scattering function can be expressed as:
\begin{equation}\label{EqS1phinc4}
  S(\mathbf{Q},\omega)^{1\,\mathrm{ph}}_{\mathrm{inc}}= \sum_d\frac{(\sigma_{\mathrm{inc}})_d}{4\pi}\frac{|\mathbf{Q}|^2}{3m_d}\exp(-2W_d)\frac{g_d(\omega)}{2\omega}\left[n(\omega)+1\right],
\end{equation}
where the partial density of states:
\begin{equation}\label{Eqpds}
  g_d(\omega)=g(\omega)\langle|\mathbf{e}_{ds}|^2\rangle,
\end{equation}
gives how much the motion of atom $d$ contributes to the density of states. The incoherent approximation is then used to interpret the one-phonon scattering from powder samples.

In the incoherent approximation, the correlations between the atom motions are neglected, and the scattering from each atom is treated as incoherent with scattering amplitude $b^{coh}_k$. This is strictly only valid in the limit of large momentum transfer $|\mathbf{Q}|$, where $|\mathbf{Q}|\gg\frac{2\pi}{d}$, and $d$ is the nearest neighbour distance in the structure. However, it is often the case that it is possible to use a wider range of $|\mathbf{Q}|$, making the calculation viable for lower incident energies, which have no access to such values of $|\mathbf{Q}|\gg\frac{2\pi}{d}$ \cite{Schober}. After experimenting with the effect of modifying the $|\mathbf{Q}|$ range, we applied the Mantid algorithm to data in the range $Q_\mathrm{max}/2<Q<Q_\mathrm{max}$.

In a sample containing more than one atomic species, the powder INS measurements do not yield the total phonon density of states, but instead give the average over the partial density of states weighted by species-dependent factors. Therefore we calculate the generalised phonon density of states:
\begin{equation}\label{EqGDOS}
  G(\omega)=\sum_d\frac{\sigma_d}{4\pi}\frac{1}{m_d}\exp(-2W_d)g_d(\omega)
\end{equation}
from our DFT model, which can be compared with our measured data according to:
\begin{equation}\label{EqGDOSexp}
  G(\omega)\propto\frac{\omega}{|\mathbf{Q}|^2}\frac{1}{n(\omega)+1}S(|\mathbf{Q}|,\omega)^{1 \mathrm{ph}}.
\end{equation}

\bibliographystyle{angew}
\bibliography{AS}